\documentclass{aastex7}

\usepackage[T1]{fontenc}
\usepackage[normalem]{ulem}    
\def\simgr{\,\hbox{\hbox{$ > $}\kern -0.8em \lower 1.0ex\hbox{$\sim$}}\,}
\def\simle{\,\hbox{\hbox{$ < $}\kern -0.8em \lower 1.0ex\hbox{$\sim$}}\,}

\shortauthors{THORSTENSEN ET Al.}
\shorttitle{FIVE SOUTHERN CATACLYSMIC BINARIES}

\accepted{2025 November 14; to appear in {\it Astronomical Journal}.}

\begin{document}
\title{Observations of Five Southern-Hemisphere Cataclysmic Binary Stars}

\author[0000-0002-4964-4144]{John R. Thorstensen}
\email{John.R.Thorstensen@dartmouth.edu}
\affil{Department of Physics and Astronomy,
6127 Wilder Laboratory, Dartmouth College,
Hanover, NH 03755-3528}

\author[0009-0005-2363-9274]{Annabelle E. Niblett}
\email{annabelle.niblett@gmail.com}
\affil{Department of Physics and Astronomy,
6127 Wilder Laboratory, Dartmouth College,
Hanover, NH 03755-3528}

\author[0000-0002-3380-8634]{Shreya Gandhi}
\email{shreyagandhi02@gmail.com}
\affil{Department of Physics and Astronomy,
6127 Wilder Laboratory, Dartmouth College,
Hanover, NH 03755-3528}

\author[0009-0005-4311-1215]{Lauren P. Zanarini}
\email{laurenzanarini@icloud.com}
\affil{Department of Physics and Astronomy,
6127 Wilder Laboratory, Dartmouth College,
Hanover, NH 03755-3528}

\author[0009-0001-6691-4863]{Gavin D. Goss}
\email{gdgoss12@hotmail.com}
\affil{Department of Physics and Astronomy,
6127 Wilder Laboratory, Dartmouth College,
Hanover, NH 03755-3528}

\author[0009-0006-6520-4495]{Arnav Singh}
\email{arnav.singh.26@dartmouth.edu}
\affil{Department of Physics and Astronomy,
6127 Wilder Laboratory, Dartmouth College,
Hanover, NH 03755-3528}

\author[0009-0005-4221-049X]{Divik Verma}
\email{divikv123@gmail.com}
\affil{Department of Physics and Astronomy,
6127 Wilder Laboratory, Dartmouth College,
Hanover, NH 03755-3528}

\author[0000-0003-1468-9526]{Ryan C. Hickox}
\email{ryan.c.hickox@dartmouth.edu}
\affil{Department of Physics and Astronomy,
6127 Wilder Laboratory, Dartmouth College,
Hanover, NH 03755-3528}

\author[0009-0004-9516-9593]{Emmanuel A. Durodola}
\email{emmanuel.a.durodola.gr@dartmouth.edu}
\affil{Department of Physics and Astronomy,
6127 Wilder Laboratory, Dartmouth College,
Hanover, NH 03755-3528}

\author[0009-0006-5300-2976]{Jiaqi Martin Ying} 
\email{martin.ying.gr@dartmouth.edu}
\affil{Department of Physics and Astronomy,
6127 Wilder Laboratory, Dartmouth College,
Hanover, NH 03755-3528}

\begin{abstract}
We present observations and analyses of five little-studied cataclysmic
binary stars in the southern celestial hemisphere.  Our new observations
are from the South African Astronomical Observatory.

The objects and salient results are as follows:
(i) 6dF0752-54 is a dwarf nova with an orbital period 
$P_{\rm orb}$ = 5.05 hr that shows a contribution from a mid-M type secondary
in its mean spectrum.  (ii) J0916-26 had been suspected
of being a magnetic CV with an eclipse
period of 3.37 hr. Our spectrum corroborates this
classification. (iii) GSC 08944 is a novalike variable with 
$P_{\rm orb}$ = 3.80 hr. Archival photometry also shows
a persistent photometric period near 4.03 hr, 
apparently from a positive superhump.  Its emission line
behavior is consistent
with an SW Sextantis-type novalike. (iv) MGAB-V253, also Gaia20eys, 
had been identified as a short-period
eclipsing novalike with $P_{\rm orb} = 1.44$ hr.  
Our spectrum shows shows broad emission lines consistent
with this, and the extensive TESS data show
a persistent modulation near 1.35 hr, evidently
a negative superhump.  It is less luminous
than most novalikes, but significantly brighter than 
quiescent dwarf novae with comparably short periods.
(v) Finally, DDE 45 shows a complicated variability history, 
cycling rapidly between high and low states for a time
and more recently showing outbursts resembling a U Gem-type 
dwarf nova. We find a 2.07 hr radial velocity period,
which also appears in archival TESS photometry.  The emission 
lines are double-peaked, with an orbital S-wave similar to 
low-inclination dwarf novae. 

\end{abstract}


\section{Introduction}

Cataclysmic binary stars, often called cataclysmic variables (CVs), are close
binary systems in which a white-dwarf primary accretes mass from a more 
extended companion (referred to as the secondary) through Roche-lobe overflow.  
The secondary usually resembles a main-sequence star, though binary evolution 
and mass
transfer can cause departures from the main-sequence mass-radius-luminosity
relation \citep{knigge06}.  The bulk of a CV's luminosity usually arises from
gravitational energy released by accretion, although the secondary star's
photosphere -- and in some cases the white dwarf's photosphere -- can 
contribute 
significant light.  In most CVs, the matter transferred from the secondary
settles into an accretion disk around the white dwarf, but if the white
dwarf's magnetic field is strong enough, the disk can be disrupted 
either partially
or entirely.  All CVs are variable stars, because the accretion processes
are never entirely steady; indeed, they are called ``cataclysmic'' because 
the variations in many systems are quite dramatic. \citet{warner95} 
describes the taxonomy of CVs and discusses their physical nature. 

Here we present observations and analyses of five southern CVs that have until
now been little-studied.  Section \ref{sec:observations} details the 
telescopes, instruments,
and procedures we used to obtain our new data.  % In [nearly] all cases 
Our analyses also incorporate publicly-available
photometric survey data (see Section \ref{sec:observations}). 

Table \ref{tab:starinfo} lists the stars we observed.  The first column
gives the short name used in this paper, and the second column
gives the primary name used in the American Association of Variable
Star Observer's International Variable Star Index (VSX)
\footnote{\url{ https://www.aavso.org/vsx/}}.
The VSX catalog lists further designations for these
objects.  All the objects here are variable stars, so the $G$ 
magnitude is only illustrative.  

\begin{deluxetable*}{lrrrrr}
\label{tab:starinfo}
\tablewidth{0pt}
\tablecolumns{6}
\tablecaption{List of Objects}
\tablehead{
\colhead{Short name} & 
\colhead{VSX Name} &
\colhead{$\alpha_{\rm ICRS}$} &
\colhead{$\delta_{\rm ICRS}$} &
\colhead{G} &
\colhead{$1/\pi_{\rm DR3}$} \\
\colhead{} &
\colhead{} &
\colhead{[h:m:s]} &
\colhead{[d:m:s]} &
\colhead{[mag]} &
\colhead{[pc]}  \\
}  % end of tablehead
\startdata
6dF0752-54 & 6dFGS g0752425-545120 & 07:52:42.46 & $-$54:51:18.9 & 17.1 & 681(25) \\ 
J0916-26 & Gaia DR2 5637827617537477504 & 09:16:04.41 & $-$26:53:52.1 & 17.6 & 528(19) \\
GSC08944 & GSC 08944-02101 & 09:19:16.27 & $-$63:07:29.8 & 13.4 & 683(6) \\
MGAB-V253\tablenotemark{a} & MGAB-V253 & 09:52:59.80 & $-$31:02:46.6 & 17.0 & 272(5) \\
DDE 45 & DDE 45 & 10:01:32.00 & $-$33:02:35.6 & 15.4 & 359(4) \\ 
\enddata
\tablecomments{The celestial coordinates, G magnitudes, and distance estimates are from
the Gaia Data Release 3 \citep{gaiamission, gaiadr3descr}.  The tabulated distances are
inverses of the parallax, without further adjustment.  
%The SIMBAD designations
%in the final column omit the Gaia numbers for the sake of space. SIMBAD
%entries for these objects can be found using their coordinates.
}
\tablenotetext{a}{Also designated Gaia20eys.}
\end{deluxetable*}

{\it Organization:}  In Sec.~\ref{sec:observations}, we describe our observations, reductions,
and use of archival data.  We then discuss the objects in order of
right ascension.  
% Table \ref{tab:summary} summarizes some of the results. 

\section{Observations and Reductions}
\label{sec:observations} 

Our observations are from the South African Astronomical Observatory near
Sutherland, South Africa.  Nearly all were obtained in February of 2025,
during a 2-week observing run under the auspices of the Dartmouth Foreign Study
Program in Astronomy. Four spectra of GSC08944 were taken in
2023 February, using the same equipment and procedures as 2025 (see below).

\subsection{Spectroscopy and Radial Velocities}

The spectra are from the SpUpNIC spectrograph \citep{spupnic} mounted 
on the 1.9-m Radcliffe
reflector.  Grating 6 yielded a dispersion of 1.34 \AA\ per pixel,
and coverage from 4100 to 6800 \AA.
A 1\arcsec.34 projected slit 
gave a resolution near 4.5 \AA\ full-width at half-maximum (FWHM).
We found and traced the stellar spectra on the 2-dimensional image
using a python-language program modeled on the IRAF {\tt apfind}
and {\tt aptrace} tasks, and then 
extracted the source and background spectra from the two-dimensional
images using a python implementation of the optimal-extraction algorithm described by \citet{horne86}.
We took spectra of a CuAr arc after every 
pointing and at least hourly as the telescope tracked, but did
not use these in the final wavelength calibration.
Instead, we first derived a pixel-wavelength relation from
arc lamps taken at zenith, and used this relation to find the 
apparent wavelengths of the $\lambda 5577$ and $\lambda 6300$ 
airglow lines, which were strong in all the background spectra.
We assumed the airglow lines were at the rest wavelengths 
given by \citet{osterbrock16}, and for each observed
spectrum, derived and applied a 2-parameter linear correction 
(zero point and dispersion) to the pixel-wavelength
relation derived from the arc lamps.  The wavelength
calibrations were therefore effectively simultaneous with 
the program spectra. 

Each clear night we took spectra of southern spectrophotometric standard
stars from \citet{hamuystds}, and used these with IRAF tasks to 
flux-calibrate our spectra.  The flux calibrations suffer uncertainties
from variable seeing losses at the rather
narrow slit, and transparency variations, but the standard-star
observations were generally consistent to within a few tenths of
a magnitude.

We measured emission-line radial velocities, mostly H$\alpha$,
by convolving the line profile with an antisymmetric function,
generally the derivative of a gaussian, and searching for the 
zero of that convolution \citep{sy80}.  The raw velocities were
shifted to the rest frame of the solar-system barycenter and 
the times of mid-exposure were corrected to the arrival time
at the barycenter.  Table \ref{tab:vtab} lists the radial velocities.

\startlongtable
\begin{deluxetable}{lrr}
\label{tab:vtab}
\tablewidth{0pt}
\tablecolumns{3}
\tablecaption{Emission-line Radial Velocities}
\tablehead{
\colhead{Time} & 
\colhead{Radial Velocity} &
\colhead{Uncertainty} \\
\colhead{BJD-TDB $-$ 2,400,000.0} &
\colhead{km s$^{-1}$} &
\colhead{km s$^{-1}$} \\
}
\startdata
\cutinhead{6dF 0752$-$54} 
60723.2885  &  $  112$ & $   8$ \\
60723.3024  &  $  154$ & $  11$ \\
60723.4516  &  $   64$ & $   9$ \\
60723.4655  &  $   98$ & $   8$ \\
60724.3049  &  $   88$ & $  12$ \\
60724.3188  &  $   88$ & $  10$ \\
60724.3328  &  $  159$ & $  11$ \\
60725.3245  &  $  -44$ & $  10$ \\
60725.3401  &  $   40$ & $   8$ \\
60725.3540  &  $  100$ & $   9$ \\
60725.3693  &  $   47$ & $  17$ \\
60725.3832  &  $  104$ & $  11$ \\
60728.3543  &  $  115$ & $  11$ \\
60728.3648  &  $  102$ & $  12$ \\
60728.3752  &  $  110$ & $  11$ \\
60728.3871  &  $   70$ & $   9$ \\
60728.3975  &  $   32$ & $   8$ \\
60728.4080  &  $   -4$ & $   9$ \\
60729.2898  &  $  -20$ & $   8$ \\
60729.3002  &  $  -36$ & $   8$ \\
60729.3107  &  $  -67$ & $   8$ \\
60729.5106  &  $  -37$ & $  11$ \\
60729.5211  &  $  -36$ & $  10$ \\
60729.5315  &  $  -36$ & $  12$ \\
60729.5443  &  $   58$ & $  10$ \\
60729.5548  &  $  101$ & $  10$ \\
60729.5653  &  $   80$ & $  14$ \\
60730.3013  &  $   17$ & $  10$ \\
60730.3117  &  $   43$ & $  10$ \\
60730.3222  &  $  -17$ & $   9$ \\
60731.2825  &  $  134$ & $   9$ \\
60731.2930  &  $  134$ & $  10$ \\
60731.3034  &  $  109$ & $   9$ \\
60732.2751  &  $   -5$ & $  22$ \\
\tablebreak
\cutinhead{GSC08944: mean of H$\alpha$ and H$\beta$} 
59985.4524  &    $  -48$ & $   8$ \\
59985.4594  &    $   -4$ & $   8$ \\
59985.4664  &    $  -71$ & $   7$ \\
59985.4752  &    $  -60$ & $   8$ \\
60720.3935  &    $  152$ & $   9$ \\
60720.4039  &    $   66$ & $  11$ \\
60720.4172  &    $   39$ & $   7$ \\
60720.4276  &    $   38$ & $   7$ \\
60720.4419  &    $   53$ & $   8$ \\
60720.4524  &    $  -15$ & $   8$ \\
60720.4676  &    $  -85$ & $  10$ \\
60720.4780  &    $  -31$ & $  13$ \\
60720.4927  &    $   10$ & $  15$ \\
60720.5031  &    $  -23$ & $  16$ \\
60720.5272  &    $   17$ & $  13$ \\
60720.5387  &    $  109$ & $  11$ \\
60720.5491  &    $  118$ & $  13$ \\
60720.5606  &    $  124$ & $  13$ \\
60720.5710  &    $   91$ & $  14$ \\
60720.5835  &    $   58$ & $  10$ \\
60720.5939  &    $   -4$ & $  10$ \\
60720.6125  &    $  -37$ & $   8$ \\
60720.6229  &    $  -39$ & $   8$ \\
60720.6337  &    $  -67$ & $  10$ \\
60721.4309  &    $  -44$ & $   8$ \\
60721.4425  &    $  -46$ & $   7$ \\
60721.4558  &    $   47$ & $   7$ \\
60721.4674  &    $   82$ & $   5$ \\
60721.4827  &    $  127$ & $   6$ \\
60721.4943  &    $  145$ & $   6$ \\
60723.3855  &    $  179$ & $   8$ \\
60723.3960  &    $  105$ & $   8$ \\
60723.4064  &    $  118$ & $   9$ \\
60723.4188  &    $   63$ & $   7$ \\
60723.4293  &    $   85$ & $   8$ \\
60723.5966  &    $   57$ & $   6$ \\
60723.6071  &    $   12$ & $   5$ \\
60723.6175  &    $   -7$ & $   6$ \\
60723.6290  &    $  -39$ & $   7$ \\
60724.3764  &    $   20$ & $   4$ \\
60724.3869  &    $  -16$ & $   5$ \\
60724.3974  &    $  -27$ & $   6$ \\
60724.4123  &    $  -63$ & $   7$ \\
60724.4228  &    $ -140$ & $   6$ \\
60724.4332  &    $ -169$ & $  11$ \\
60724.5530  &    $  -59$ & $   7$ \\
60724.5635  &    $  -37$ & $   7$ \\
60729.3886  &    $  108$ & $   7$ \\
60729.3991  &    $  129$ & $   6$ \\
60729.4096  &    $  104$ & $   6$ \\
60730.3533  &    $  127$ & $   5$ \\
60730.3638  &    $  109$ & $   7$ \\
60730.3742  &    $   32$ & $   9$ \\
\cutinhead{MGAB-V253: mean of H$\alpha$ and H$\beta$}
60722.4153  &    $  -22$ & $  10$ \\
60722.4258  &    $  -68$ & $  10$ \\
60722.4423  &    $   45$ & $  15$ \\
60722.4528  &    $   78$ & $  10$ \\
60722.4717  &    $   60$ & $  16$ \\
60722.4787  &    $  -60$ & $  13$ \\
60722.4857  &    $  -82$ & $  17$ \\
60722.5060  &    $   21$ & $  34$ \\
60722.5130  &    $   62$ & $  17$ \\
60722.5270  &    $    5$ & $  26$ \\
60722.5360  &    $  -27$ & $  15$ \\
60722.5499  &    $  -78$ & $  26$ \\
\cutinhead{DDE 45}  
60722.5719  &    $  219$ & $  17$ \\
60722.5824  &    $  243$ & $  23$ \\
60722.5947  &    $   47$ & $  12$ \\
60722.6052  &    $ -175$ & $  16$ \\
60722.6156  &    $ -240$ & $  23$ \\
60722.6283  &    $  104$ & $  32$ \\
60722.6387  &    $   84$ & $  20$ \\
60724.3571  &    $   27$ & $  15$ \\
60728.3134  &    $ -198$ & $  15$ \\
60728.3218  &    $  -48$ & $  24$ \\
60728.3302  &    $   37$ & $  23$ \\
60728.3386  &    $   26$ & $  16$ \\
60729.3298  &    $  107$ & $  14$ \\
60729.3382  &    $   52$ & $  10$ \\
60729.3465  &    $  -95$ & $   8$ \\
60729.3549  &    $ -196$ & $  11$ \\
60729.3647  &    $ -219$ & $  14$ \\
60729.3731  &    $   39$ & $  17$ \\
60729.4679  &    $    7$ & $  14$ \\
60729.4763  &    $  163$ & $  10$ \\
60729.4846  &    $  308$ & $   8$ \\
60729.4930  &    $  386$ & $  10$ \\
60729.5895  &    $  149$ & $  11$ \\
60729.5978  &    $   36$ & $   9$ \\
60730.4522  &    $  150$ & $  16$ \\
60731.3228  &    $   38$ & $  12$ \\
60731.3312  &    $  -39$ & $  13$ \\
60731.3396  &    $  -87$ & $  19$ \\
60731.3480  &    $   95$ & $  15$ \\
\enddata
\tablecomments{
Times given are the barycentric Julian date of mid-exposure, 
in the TDB system, minus 2,400,000.0.  Note that they differ from
Modified Julian Dates by 0.5 d.  Velocities are for H$\alpha$ if not
otherwise specified.} 
\end{deluxetable}

\subsection{New and Archival Photometry.}

\subsubsection{SAAO SHOC}

We obtained time-series photometry with the Sutherland High-Speed
Optical Camera (SHOC; \citealt{coppejans13}) mounted on the 
SAAO 1.0 m telescope.  No filters were used.  The $1024 \times 1024$-pixel
active imaging area of the CCD subtended 2.85 arcmin,
and was binned $4 \times 4$, yielding 0\arcsec.668 per effective pixel.
In most cases our individual exposures were 30 seconds. 
Because of the frame-transfer CCD, the dead time between exposures 
was only 6.76 milliseconds.

We reduced the images by subtracting an average bias and dividing
by the median of flat field images of the twilight sky.  To measure
instrumental magnitudes, we used a script that called the 
IRAF aperture photometry task {\tt phot}, usually with a 
6\arcsec diameter software aperture. The telescope occasionally suffered tracking and guiding problems, but the measurement script usually
followed the shifting pixel coordinates successfully. 
To form the time series, we differenced the instrumental magnitudes
of the target and a nearby, somewhat brighter comparison star, 
thus controlling for seeing variations and light cloud cover.  
Finally, we added the Gaia G magnitude of the reference star to the 
differential magnitudes, shifting them very approximately to G.
We caution that our unfiltered CCD passband was not
matched to the Gaia passband, so this is not precise. 

Table~\ref{tab:obsjournal} gives a journal of our SHOC observations.

\begin{deluxetable}{lrrrr}
\label{tab:obsjournal}
\tablewidth{0pt}
\tablecolumns{5}
\tablecaption{SHOC Time-Series Photometry} 
\tablehead{
\colhead{Start} & 
\colhead{$N$} & 
\colhead{$\Delta t$} &
\colhead{HA start} &
\colhead{HA end} \\
\colhead{(UT)} & 
\colhead{} &
\colhead{(s)} &
\colhead{HH:MM} & 
\colhead{HH:MM} \\
}
\startdata
\cutinhead{J0916-26}
2025-02-15 22:13 & 459 & 30 & +00:04 & +03:55 \\
2025-02-21 18:34 & 90 & 30 & $-$03:12 & $-$02:28 \\
2025-02-22 20:04 & 123 & 30 & $-$01:38 & $-$00:32 \\
2025-02-23 18:45 & 180 & 30 & $-$02:53 & $-$01:23 \\
2025-02-24 18:23 & 230 & 30 & $-$03:11 & $-$01:16 \\
2025-02-25 19:02 & 38 & 30 & $-$02:28 & $-$02:10 \\
\cutinhead{MGAB-V253}
2025-02-17 19:56 & 170 & 30 & $-$02:43 & $-$01:18 \\
2025-02-18 19:43 & 260 & 15 & $-$02:51 & $-$01:45 \\
2025-02-19 23:46 & 106 & 30 & +01:17 & +02:16 \\
2025-02-24 20:28 & 105 & 15 & $-$01:43 & $-$01:16 \\
\enddata
\tablecomments{$N$ is the number of integrations;
$\Delta t$ is the interval between successive
integration start times. The dead time was very short.
}
\end{deluxetable}

\subsubsection{Archival photometry.}

Several sources of archival photometry proved useful.

{\it TESS.} We used the {\tt lightkurve} 
package \citep{lightkurve} to download 
light curves from the Transiting Exoplanet
Survey Satellite (TESS; \citealt{rickertess}).  Since these
are nearly-unbroken segments of 28 days, they do
not suffer from the 1 cycle d$^{-1}$ aliasing problems
that plague single-site ground-based measurements.

{\it ATLAS.} We obtained light curve from the Asteroid Terrestrial-impact 
Last Alert System (ATLAS; \citealt{tonryatlas}) using
their web-based forced photometry server \citep{shinglesatlas}.
The flux this returns is the difference between the 
individual frame and an average, so (for example) 
eclipses tend to appear as negative flux values.  We read
these with a program that discarded extreme values and 
values with large estimated errors, 
and corrected the timings as described below. ATLAS 
light curves typically cover more than 5 years, with 
several thousand usable measurements. 

{\it ASAS-SN.}  \citet{asasn-descr} describe the All-Sky
Automated Survey for Supernovae (ASAS-SN). Light 
curves for objects in this survey can be generated 
using their photometry tool \citep{asassn_phot}. These
are usually not quite as dense as ATLAS light curves but
cover a similarly long time frame.

{\it Gaia.} For some targets, the Gaia DR3 \citep{gaiamission,
gaiadr3descr} includes
epoch photometry, that is, fluxes from individual observations,
which can be downloaded from the Vizier catalog server
\citep{vizier}.

\subsubsection{A Note on Timing.}

The times in our ground-based image headers are based on Coordinated
Universal Time (UTC), as is usually standard.  We corrected these
to times of mid-exposure and shifted them to the arrival time at the solar
system barycenter to account for varying position of the
Earth in its orbit.  ATLAS data are referred to geocentric
UTC, so we applied similar barycentric time-of-flight corrections to 
those.  ASAS-SN times are already barycentric.
The barycentric corrections left the data 
in the UTC system, which is tied to within 1 second
to the phase of the earth's rotation, and hence 
is not uniform.

The TESS observations are referred to TDB (Barycentric Dynamical 
Time), which is uniform.  
At the present time, TDB can be obtained from UTC by
adding approximately 69 seconds. For consistency with TESS,
we adjusted all our UTC-based barycentric times to TDB.

Timings from Gaia are 
originally in yet another system, TCB, which runs at the
rate of a clock outside the sun's gravitational influence,
and gains nearly a half-second per year on TDB.  We adjusted
Gaia timings to TDB using the {\tt astropy.time} package.

\section{Results for Individual Objects}

\subsection{6dF 0752-54} 

Apparently this star was first identified as a CV by \citet{mahony10},
in spectra from the UK Schmidt 6dFGS quasar survey.  The ASAS-SN 
light curve shows an active dwarf nova with quiescence near 
$g = 17$ and outbursts to $\sim 13$ mag.
The outbursts are typically separated
by $\sim 100$ d, with durations of $\sim 10$ d.  Their profiles show
rapid rises and a more gradual decline, as is commonly seen in 
U Gem-type dwarf novae.

\begin{figure}
\vspace*{-1cm}\hspace*{-0.0cm}\includegraphics[width=6.5 truein]{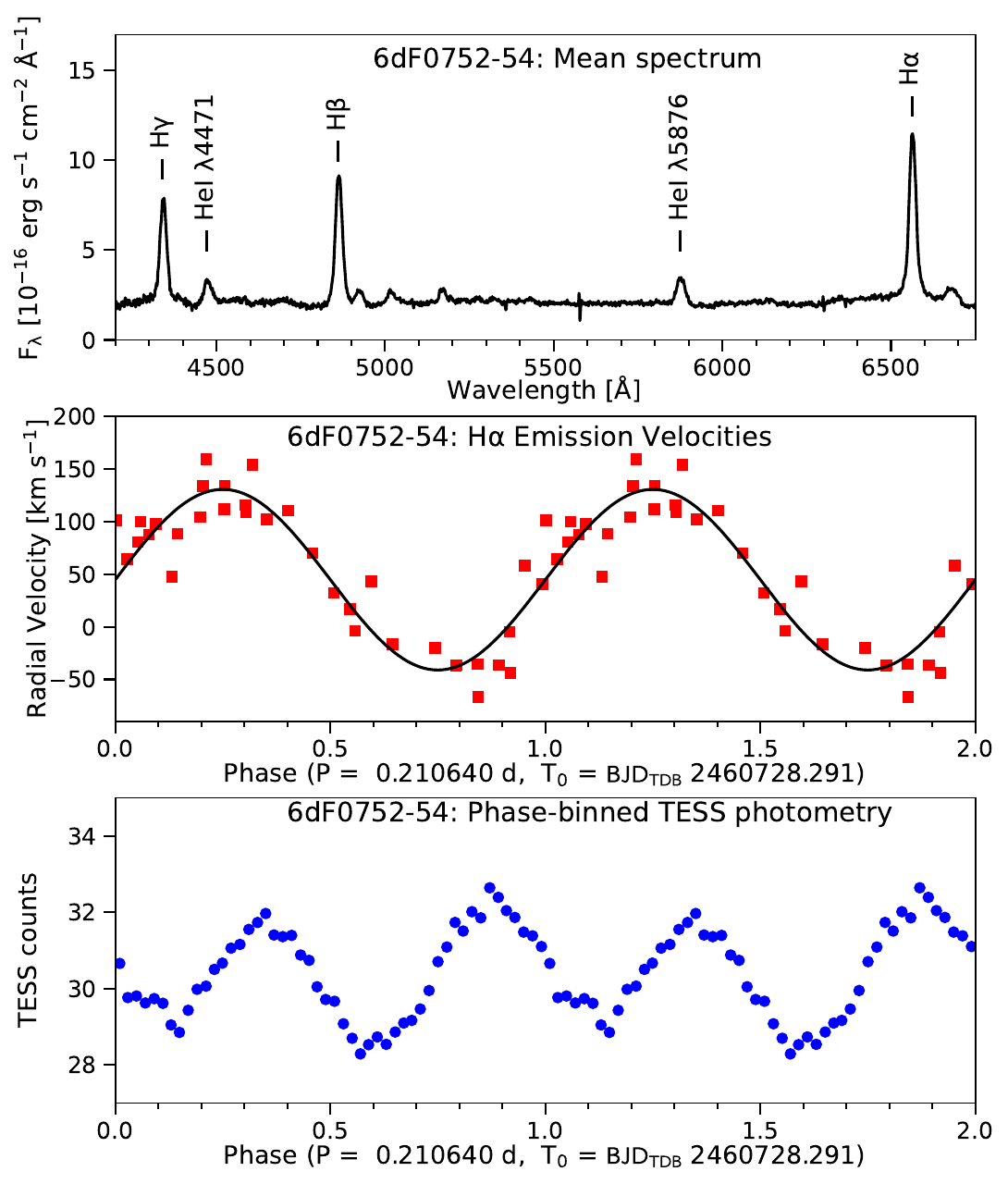}
\caption{
{\it Upper:} Mean fluxed spectrum of 6dF 0752-54. 
{\it Middle:} H$\alpha$ radial velocities from 2025 February, folded
on the orbital period, together with the best-fitting sinusoid.
{\it Lower:} TESS photometry of 6dF 0752-54 averaged into bins
on the orbital period.  
}
\label{fig:6dfspecvplot} 
\end{figure}

The mean spectrum (Fig.~\ref{fig:6dfspecvplot}, top) shows the strong
Balmer and HeI emission lines typical of a quiescent dwarf nova.
Toward the red end, a contribution from a late-type star is 
discernible.  Lacking coverage farther to the red, we were unable 
to determine a precise spectral type, but we were able to 
cancel the late-type features reasonably well by subtracting
away a spectrum of an M4-type dwarf scaled to around
25 per cent of the continuum near 6500 \AA .

The H$\alpha$ radial velocities showed an unambiguous 
periodicity at 0.2106(3) d, or $\sim$ 5.05 h, in the usual
range of U Gem-type orbital periods. The middle panel of
Fig.~\ref{fig:6dfspecvplot} shows these folded on our 
adopted ephemeris.
TESS observed the 
object in 2023, in Sectors 61, 62, and 63.  A Lomb-Scargle
periodogram of the concatenated TESS-SPOC light curves shows
a strong, unique peak at a period near 0.10533 d, half the 
orbital period. The modulation is subtle in the folded 
light curve, but phase-binning the data on twice the 
TESS period reveals a clear
double-peaked modulation on $P_{\rm orb}$, with a 
shape consistent with ellipsoidal variation
(Fig. \ref{fig:6dfspecvplot}; lower panel). The amplitude is 
relatively low, as might be expected given the 
secondary star's modest contribution to the spectrum.  
The periodogram
of the ATLAS light curve (with outbursts omitted) also 
shows a non-unique peak near $P_{\rm orb} / 2$, 
consistent with the ellipsoidal variation.
Because ATLAS observed over many seasons, 
this should give a more accurate period than the TESS or velocity data alone.
The ATLAS variation is just visible in folded data.
Examining the folded data while gradually 
varying the period refines $P_{\rm orb}$ to
$P_{\rm orb} = 0.210640(6)$ d, where the
error is a conservative by-eye estimate.  The interval
between the TESS and spectroscopic data is nearly 
800 days; in that interval, the accumulated uncertainty
in the phase computed from 
the ATLAS period is $\sim 0.12$ cycle.

The phase in Fig.~\ref{fig:6dfspecvplot}
is based on the spectroscopic epoch and the nominal
period.  If the emission-line velocities trace the 
compact object, then it corresponds to quadrature phase,
which in turn should correspond to a maximum in the ellipsoidal 
variation.  Unfortunately, the accumulated phase uncertainty
between the TESS and velocity observations precludes an
accurate phase comparison.

%   vemn_Ha.tdb with period fixed at atlas ellipsoidal
%     Epoch plus     4.060608e-03 gives    6.010168e+02
%     Epoch minus    4.064837e-03 gives    6.010153e+02
%Mean error bar =    4.062723e-03
%     K-Vel plus     1.051688e+01 gives    6.010155e+02
%     K-Vel minus    1.051688e+01 gives    6.010155e+02
%Mean error bar =    1.051688e+01
%     Gamma plus     7.391737e+00 gives    6.010155e+02
%     Gamma minus    7.391737e+00 gives    6.010155e+02
%Mean error bar =    7.391737e+00
%FISH subcommand:pr
%auto_measure_dgau_12.00_05.00_6562.817_corrected_ut_to_tdb
%   Epoch = 6.0728290481e+04
% Delta    Epoch -> 1.0000000000e+00
%     Per = 2.1064000000e-01
% Delta      Per -> 0.0000000000e+00
%   K-Vel = 8.5802204974e+01
% Delta    K-Vel -> 1.0000000000e+00
%   Gamma = 4.4677766695e+01
% Delta    Gamma -> 1.0000000000e+00
%5.4637770382e+02   n = 34
%mean epoch = 6.0727815885e+04
%For four-param fit, err of single measure = 24.884295
%
%
%% 2460728.2892(4) E + 0.210640(6) 
% 2460728.2905(4) E = 0.210640(6) ... Ha_vemn.tdb

%
%\begin{figure}
%\vspace*{-1cm}\hspace*{-0.0cm}\includegraphics[width=7.0 truein]{6dftessellipsoidal.pdf}
%\caption{
%{\it Upper:} Mean fluxed spectrum of 6dF 0752-54. 
%{\it Lower:} H$\alpha$ radial velocities from 2025 February, folded
%on the orbital period, together with the best-fitting sinusoid.
%}
%\label{fig:6dftessellipsoidal} 
% \end{figure}

\subsection{J0916-26}   % or whatever

This was discovered in a search for variable
stars in the Gaia DR2 data by \citet{gaiadr2vars},
who discovered an eclipse with a period near 0.1405 
d and suggested its out-of-eclipse variation resembled
an AM CVn star (double-degenerate CV), though its
orbital period was longer than AM CVn stars.
Denis Denisenko drew attention to their work
in a post to the 
Facebook group ``Cataclysmic Variables'', on 2018
November 24. He used archival Catalina
Real Time Survey (Siding Spring Survey) to refine the period to 
0.140497 d, and suggested that \citet{gaiadr2vars} had
meant to suggest an AM Her (or polar) classification rather
than AM CVn.  He noted that the out-of-eclipse variation
appeared smooth, similar to an EW binary (near-contact
eclipsing binaries with ellipsoidal variation; 
\citealt{gcvs17}), but noted the star's proximity
to an X-ray source, 1RXS J091605.8-265403.  The VSX 
lists its type as ``AM+E'', that is, an eclipsing AM 
Her star.  

\begin{figure}
\vspace*{-1.5cm}\hspace*{0.0cm}\includegraphics[width=6.5 truein]{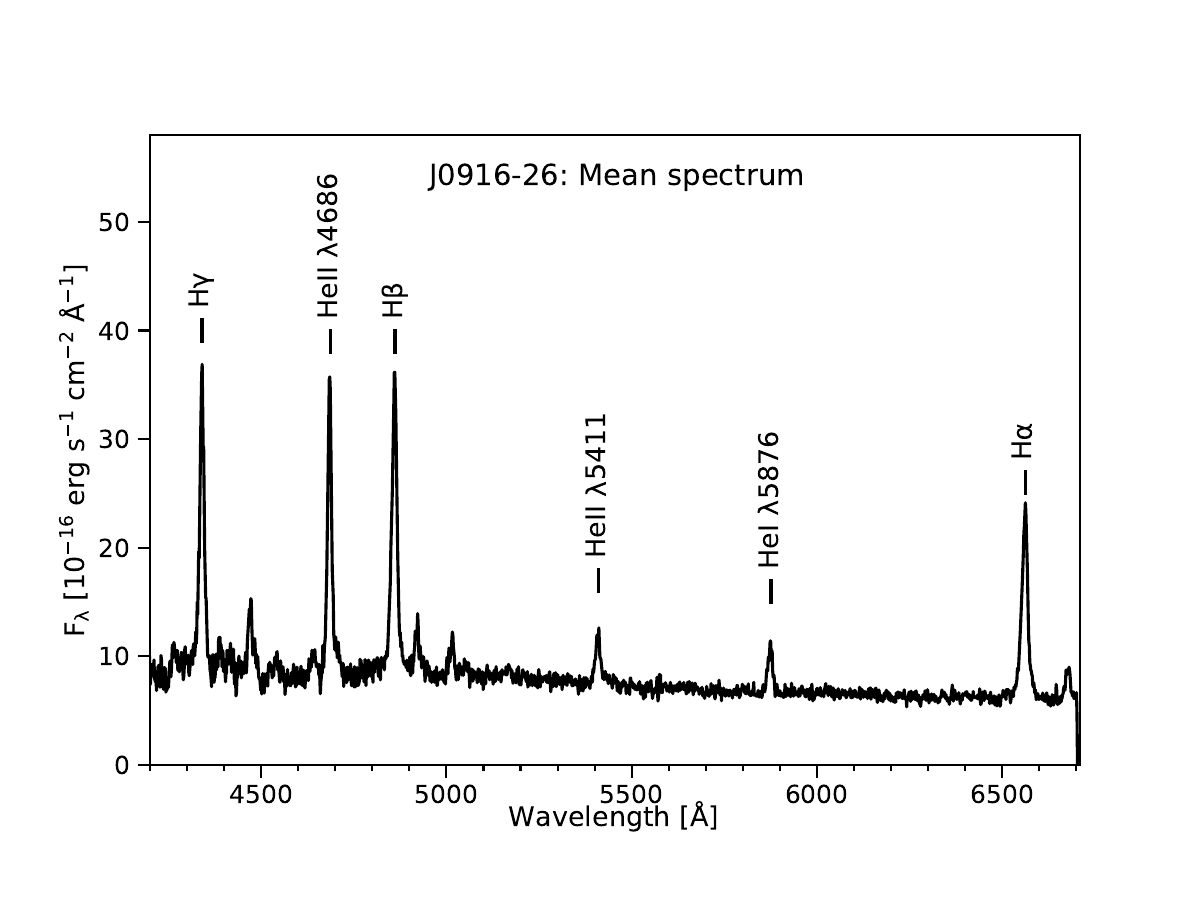}
\caption{Mean of six 1000-s spectra of J0916-26 taken 2025 Feb.~15,
with some prominent emission features labeled.
}
\label{fig:j0916spec} 
\end{figure}

Fig.~\ref{fig:j0916spec} shows the average of six 
spectra, each exposed for 1000 sec, taken
2025 February 15. 
Note the strong emission at HeII $\lambda 4686$ 
and in the Balmer and HeI lines.
HeII $\lambda 5411$ is also easily visible.  CVs with 
$\lambda 4686$ comparable to H$\beta$ are nearly all AM Her stars, 
supporting the VSX classification.
Because the likely orbital period was known, we did not 
take pains to independently determine a period from 
radial velocities, so our spectroscopic coverage is relatively
sparse.

\begin{figure}
\vspace*{-1.5cm}\hspace*{0.0cm}\includegraphics[width=6.5 truein]{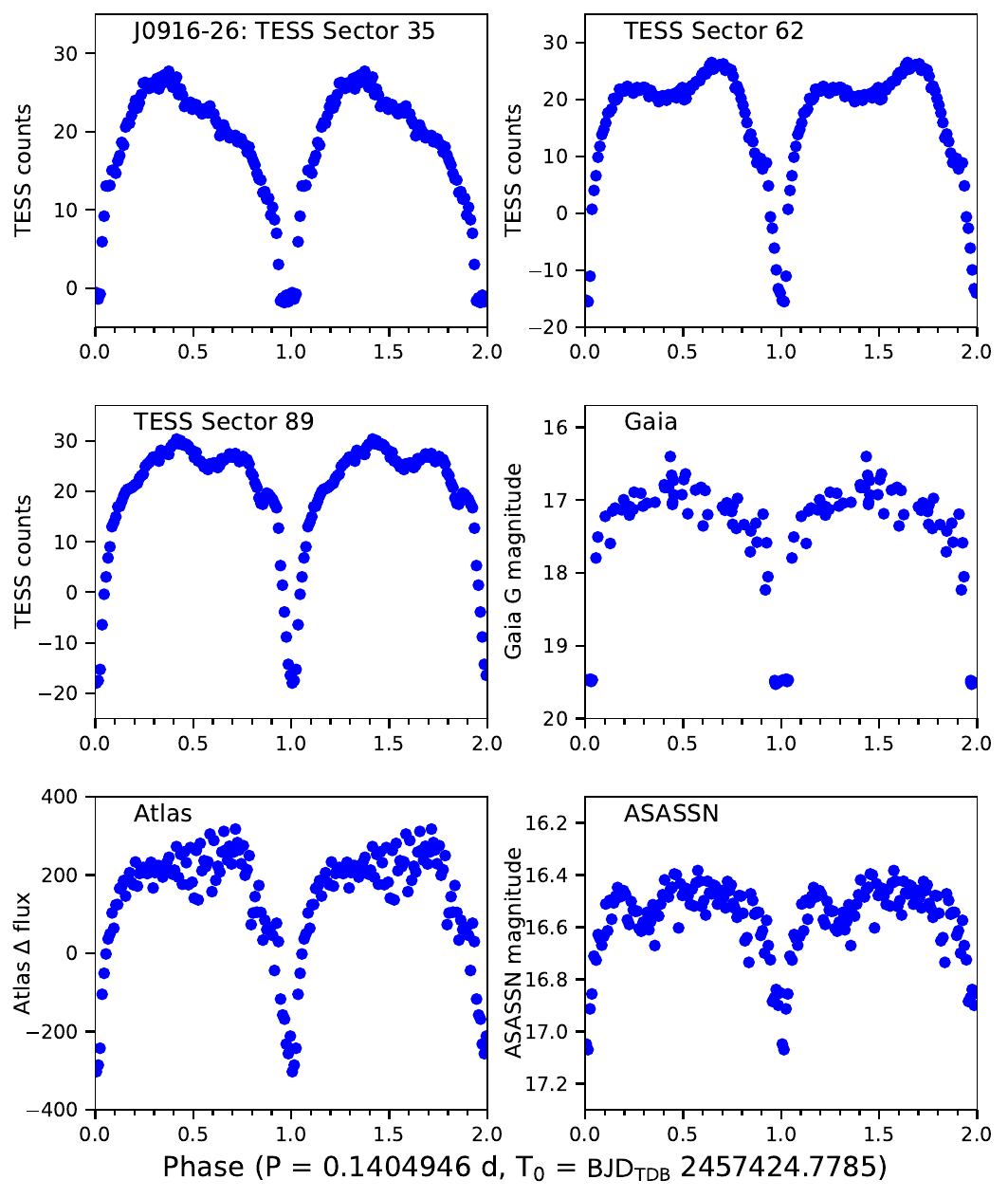}
\caption{Archival photometry of J0916$-$26 folded on our adopted
ephemeris. The Gaia points are single observations, while all
the others are averaged into 100 phase bins per cycle.
}
\label{fig:j0916_binned_archival} 
\end{figure}

Fig.~\ref{fig:j0916_binned_archival} shows archival 
photometry.  We searched the archival data for periods and 
confirmed Denisenko's period, which we then refined
slightly by tabulating the eclipse epochs from the 
various sources and fitting them with a linear ephemeris,
$$t = {\rm BJD_{TDB}}\ 2457424.7785(17) + 0.1404946(1),$$ 
where $t$ is the time of mid-eclipse.
Except for the rather sparse Gaia data, the light
curves in Fig.~\ref{fig:j0916_binned_archival} are phase-binnned.  
The TESS light curves in particular show significant 
variations from epoch to epoch.  

\begin{figure}
\vspace*{-1.5cm}\hspace*{0.0cm}\includegraphics[width=6.5 truein]{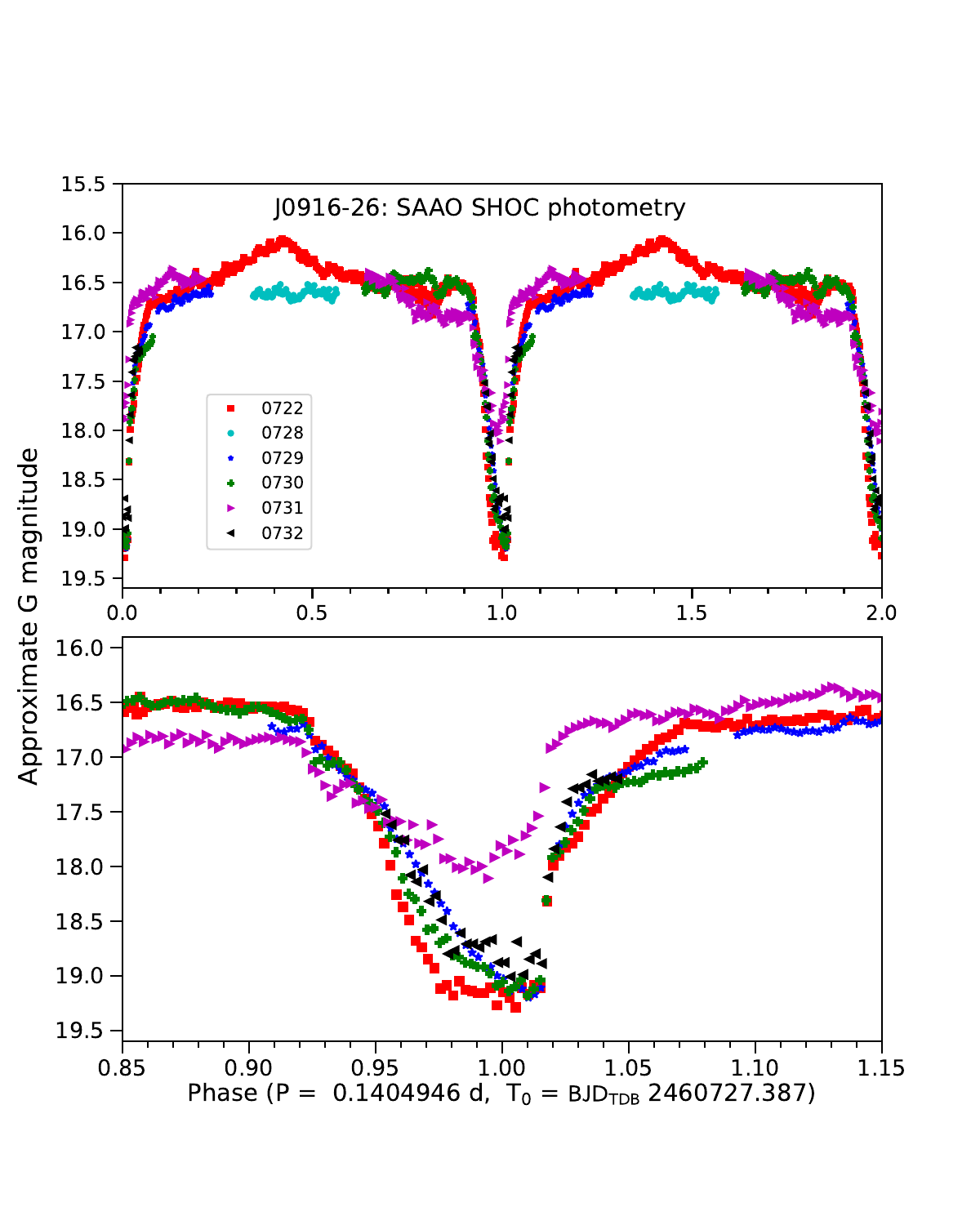}
\caption{Time series photometry of J0916$-$26 from the SAAO SHOC.
The different symbols are from different nights, keyed by the last
four digits of the Julian date. The lower panel magnifies the phases
near eclipse.
}
\label{fig:j0916_shoc} 
\end{figure}

Our SAAO time-series photometry is shown in 
Fig.~\ref{fig:j0916_shoc}. The light curve is variable
from night-to-night.  At this epoch, the eclipse ingress was
relatively gradual, but all the light curves show a sharp 
jump in egress, near phase 0.02 in our adopted ephemeris. 

\subsection{GSC08944} 

The VSX credits Marius Bajer with the discovery of 
this variable star; he posted a light curve
\footnote{\url{https://www.aavso.org/vsx_docs/684152/3635/GSC\%2008944-02101\%20HJD\%20plot.png}}
from ASAS-SN showing its variation over several seasons,
and classified it as a likely novalike variable.
It is classified as a hot subdwarf candidate
in SIMBAD.  

\begin{figure}
\vspace*{-1cm}\hspace*{-0.0cm}\includegraphics[width=6.5truein]{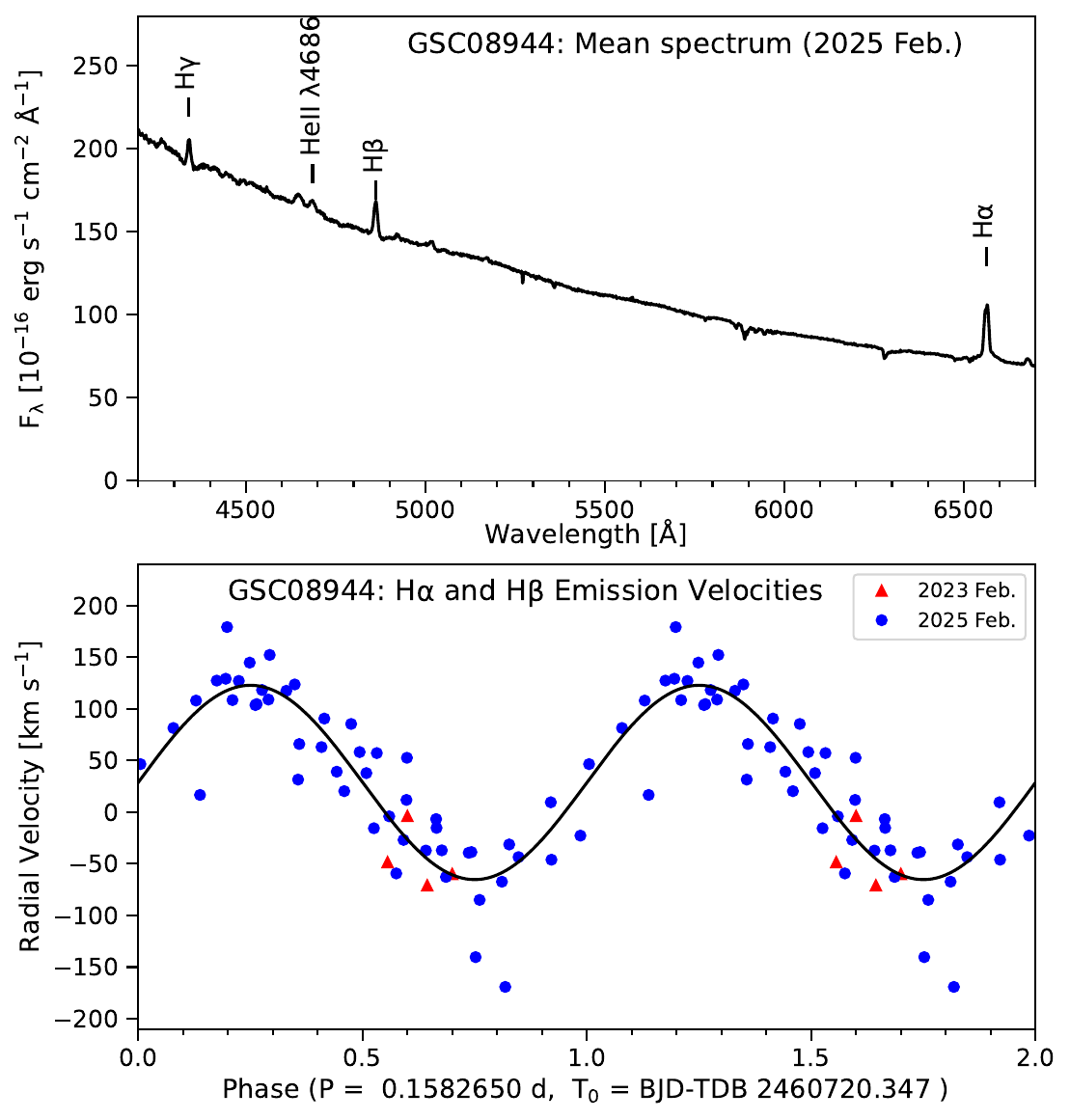}
\caption{
{\it Upper:} The mean fluxed spectrum of GSC08944, from 2025 February.
{\it Lower:} Weighted-average radial velocities of the H$\alpha$ and H$\beta$
emission lines, folded on the adopted orbital period, with the best-fitting
sinusoid.
}
\label{fig:gsc08944specfold} 
\end{figure}

We obtained four consecutive 600 s exposures
2023 February 09. In 2025 February we obtained 
49 exposures on six nights, totaling over 
12 hours and spanning 6.18 hours of hour angle.
The mean spectrum (Fig.~\ref{fig:gsc08944specfold})
shows a strong blue continuum together with 
emission lines consistent with a novalike
variable.  The H$\alpha$ emission 
equivalent width is $\sim 8$ \AA\ and its FWHM is near 
15 \AA.  There is also weak, broad emission
from HeII $\lambda$ 4686 and the high-excitation
blend near $\lambda$ 4640.  

\begin{figure}
\vspace*{-1cm}\hspace*{-0.0cm}\includegraphics[width=6.5 truein]{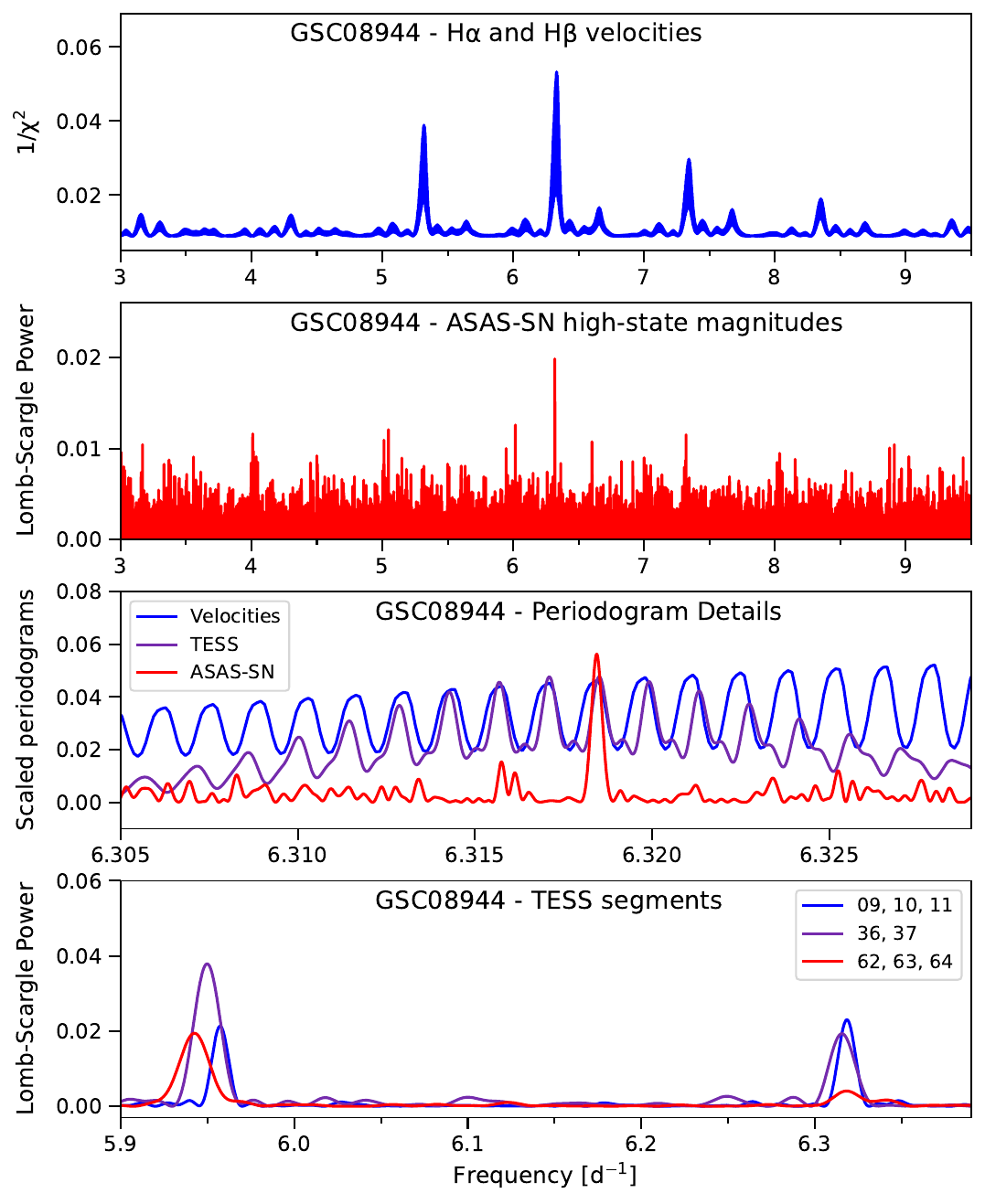}
\caption{
{\it Upper:} Period search of the weighted average of the H$\alpha$ and H$\beta$
emission line wing velocities.  {\it Upper middle:} Lomb-Scargle periodogram
of the ASAS-SN magnitudes, restricted to magnitudes between 13.2 and 13.8
(omitting low states).  {\it Lower middle:} Close-up view of the periodograms
near the adopted $P_{\rm orb}$, also including that of the combined
TESS data.  The ASAS-SN and TESS periograms are scaled for visibility.
{\it Bottom:} Periodograms of the three near-contiguous TESS data sets, 
showing the regions around the modulations we identify as $P_{\rm orb}$ 
(to the right) and $P_{\rm sh}$.
}
\label{fig:gsc08944periodograms}
\end{figure}

We first measured radial velocities using the 
derivative of a gaussian as the convolution 
function, but obtained better results using 
two gaussians separated by slightly more than
the line width. We measured H$\alpha$ and
H$\beta$. A period search 
of the H$\beta$
velocities selected a period near 0.158 d 
(6.33 c d$^{-1}$);
the H$\alpha$ velocities corroborated this, 
but also a showed period near 0.188 d, which
differs in frequency by one cycle per day and is likely an alias of the 0.158-d period.
The top panel of Fig.~\ref{fig:gsc08944periodograms}
shows a period search of the weighted average H$\alpha$
and H$\beta$ radial velocities.

TESS has covered this star extensively,
in sectors 9, 10, and 11 (2019), 
36 and 37 (2021), and 62, 63, and 64 (2023). 
% [ PUT THESE IN A TABLE? ]. 
We formed three time series by concatenating the
data from adjacent TESS sectors and 
searched them for periodicities using
the Lomb-Scargle algorithm.  All three sectors
showed a strong periodicity near 5.95 c d$^{-1}$,
as well as a varying-strength periodicity
near 6.316 c d$^{-1}$, consistent with the 
radial-velocity period.  We also concatenated
% Stolz-Schoembs quantity (Psh - Porb) / Porb
% is 0.0615.  Updates Stolz-Schoembs relation
% from Bruch 2023MNRAS.525.1953B gives 0.085,
% but there's huge scatter.
all the TESS data\footnote{The data from 
sectors 9 through 11 were from the {Quick-Look Pipeline} and 
were normalized to the mean flux; we multiplied
them by 650 to roughly match the other sets 
before combining sets.} and searched the
combined sets, which split the periodicities
into aliases spaced by about 1 cycle per 
2 years.  The 2-year gap in the radial-velocity
data led to a similar fine-scale aliasing.
The ASAS-SN photometry, with the lower-state (magnitude
> 13.6) edited out, showed a significant peak
consistent with the strongest 0.158-d radial-velocity 
period (upper middle panel of Fig.~\ref{fig:gsc08944periodograms}).
The lower middle panel of Fig.~\ref{fig:gsc08944periodograms} shows
that the ASAS-SN periodicity selects a single
alias in the radial velocity and TESS data, 
near 6.31845(7) c d$^{-1}$,
where the uncertainty is a conservative estimate,
corresponding to 0.158265(2) d.  We adopt this as
$P_{\rm orb}$; the lower panel of Fig.~\ref{fig:gsc08944specfold} shows
the emission radial velocities folded at this period. 

\begin{figure}
\vspace*{-1cm}\hspace*{-0.0cm}\includegraphics[width=6.5truein]{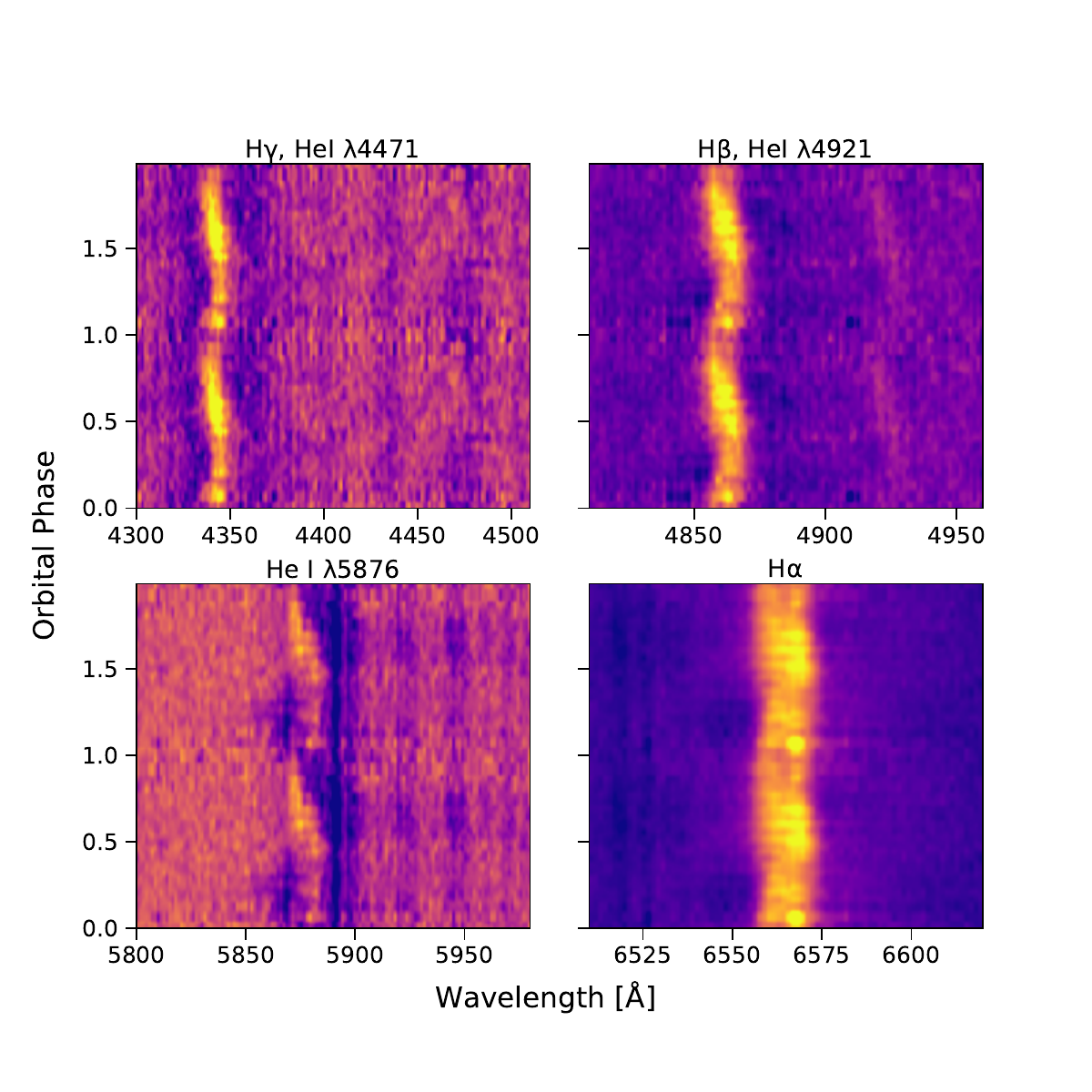}
\caption{
Portions of the phase-resolved average spectrum of GSC08944 near spectral lines of
interest.
}
\label{fig:gsc08944trailed} 
\end{figure}

To explore how the emission lines behave with orbital phase,
we formed phase-resolved spectrograms from the 2025 data
by (1) normalizing the spectra to the continuum, (2) computing
the orbital phases at which the spectra were taken, (3) 
computing weighted mean average spectra centered on 
each of 100 phase bins, \sout{the weights being from a truncated
Gaussian in phase,} and (4) stacking these spectra into a
two-dimensional image.  Fig.~\ref{fig:gsc08944trailed} shows
the results for several spectral lines. The velocity
modulation is especially clear in H$\beta$, H$\gamma$,
and HeI $\lambda 4921$; H$\gamma$ shows weak absorption
wings as well.  H$\alpha$ has a more complicated
profile, nearly two-peaked at some phases.  HeI $\lambda$5876
has remarkable behavior, with apparent absorption interleaving
with the emission.  The prominent sharp absorption is 
evidently from the interstellar D lines ($\lambda\lambda
5889$ and $5895$).  The diffuse blueward absorption
between phases 0.1 and 0.4 or so is echoed in the 
HeI$\lambda$ 4471 line, though that shows almost no
emission.  The phase-dependent HeI absorption is reminiscent of 
that seen in the prototypical SW Sex-type novalike
variable, PX And \citep{thorpxand91}.

\begin{figure}
\vspace*{-1cm}\hspace*{-0.0cm}\includegraphics[width=7.0 truein]{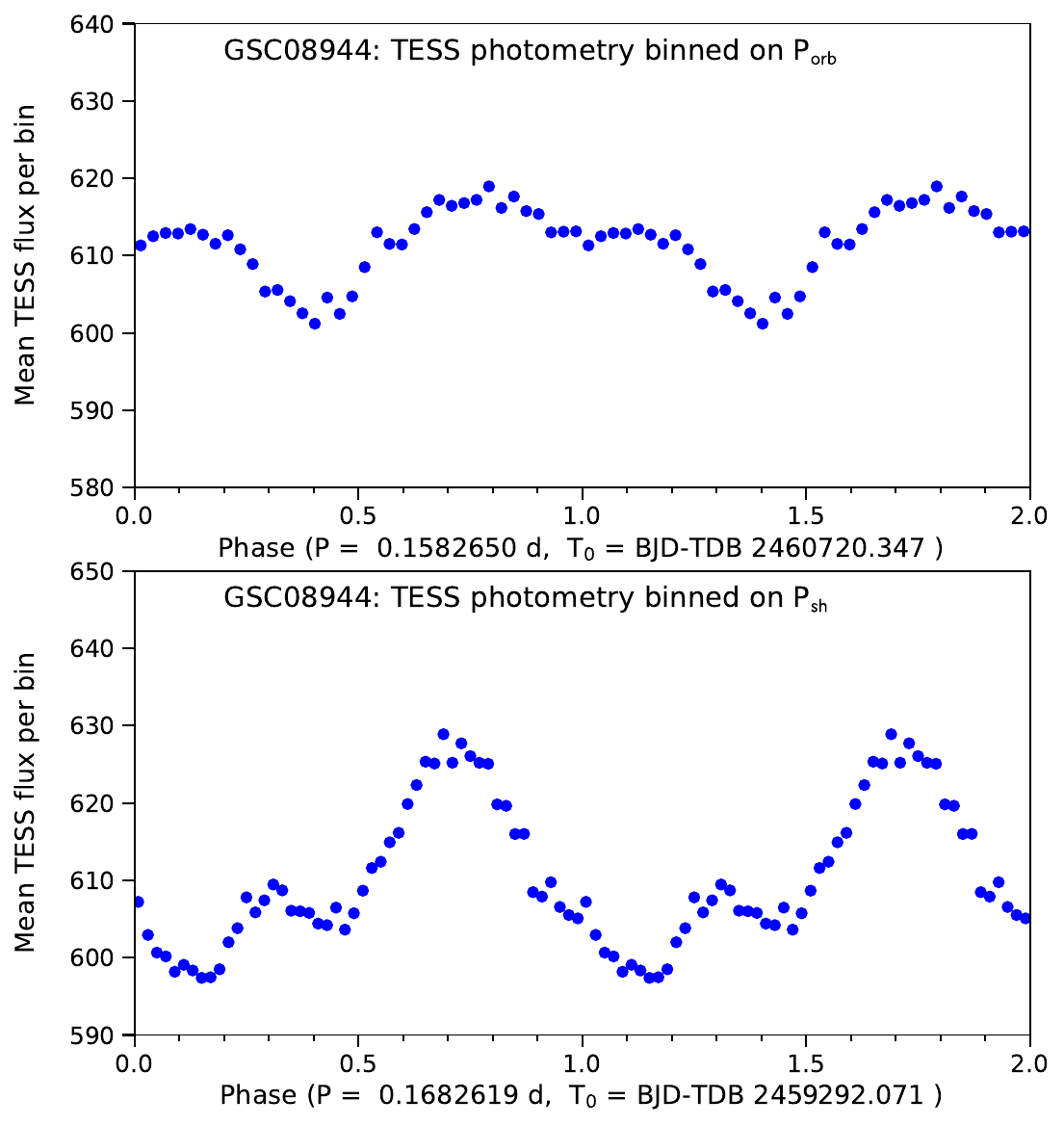}
\caption{
All the TESS data summed into phase bins on $P_{\rm orb}$ (upper) and
$P_{\rm sh}$ (lower).  The period chosen for the fold on $P_{\rm sh}$
from an arbitrary peak in the aliased periodogram and the underlying
modulation is unlikely to be coherent over long time scales.
}
\label{fig:gsc08944lightcurves} 
\end{figure}

As noted earlier, the TESS light curves show persistent
periodicity near 5.95 cycle d$^{-1}$ (0.168 d) in addition 
to the orbital period near 0.158 d.  The lower panel of
Fig.~\ref{fig:gsc08944periodograms} shows periodograms
of the three near-contiguous TESS observations in the
range covering these features.  The 5.95 cycle d$^{-1}$
feature appears to wander slightly in frequency, while
the 6.13 cycle d$^{-1}$ feature remains steady.  This is
expected if the longer-period signal is a {\it positive
superhump}, a persistent modulation at a period $P_{\rm sh}$ 
longer than $P_{\rm orb}$.  Positive superhumps are usually 
attributed to the apsidal 
precession of an elliptical accretion disk.  \citet{stolzschoembs84}
noticed that the superhump period excess, 
$\epsilon = (P_{\rm sh} - P_{\rm orb}) / P_{\rm orb}$, correlates with 
$P_{\rm orb}$ among SU-UMa type dwarf novae, and a version 
of their relationship applies to novalike variables as well.
\citet{bruch23b} recently analyzed TESS 
data for a sizeable sample of novalike variables, and 
updated the Stolz-Schoembs relation.  Assuming we have identified
the periods correctly, GSC 08944 has $\epsilon ~ \sim 0.061$, 
a bit smaller than expected from the updated relationship,
but well within the spread of the relation.  Fig.~\ref{fig:gsc08944lightcurves}
shows all the TESS data folded on $P_{\rm orb}$ and on one of the many possible alias periods
near $P_{\rm sh}$. Both modulations are clearly seen despite their modest amplitudes. The superhump modulation
is unlikely to be coherent over long time scales.

\subsection{MGAB V253 = Gaia20eys}

Gabriel Murawski found eclipses in this 
object\footnote{\url{https://www.aavso.org/vsx_docs/689866/3120/148.249173\%20-31.046281\%20\%28E\%29.png}}
with a period near 0.0598736 d, and classified it as an eclipsing
novalike variable.  It later appears in the Gaia alerts index as Gaia20eys. 
%The Gaia light curve shows the $G$ magnitude varying mostly
%from around 16 to around 18, with isolated faint points characteristic
%of an eclipse, and no dwarf nova outbursts.   

\begin{figure} 
\vspace*{-1cm}\hspace*{-0.0cm}\includegraphics[width=6.5 truein]{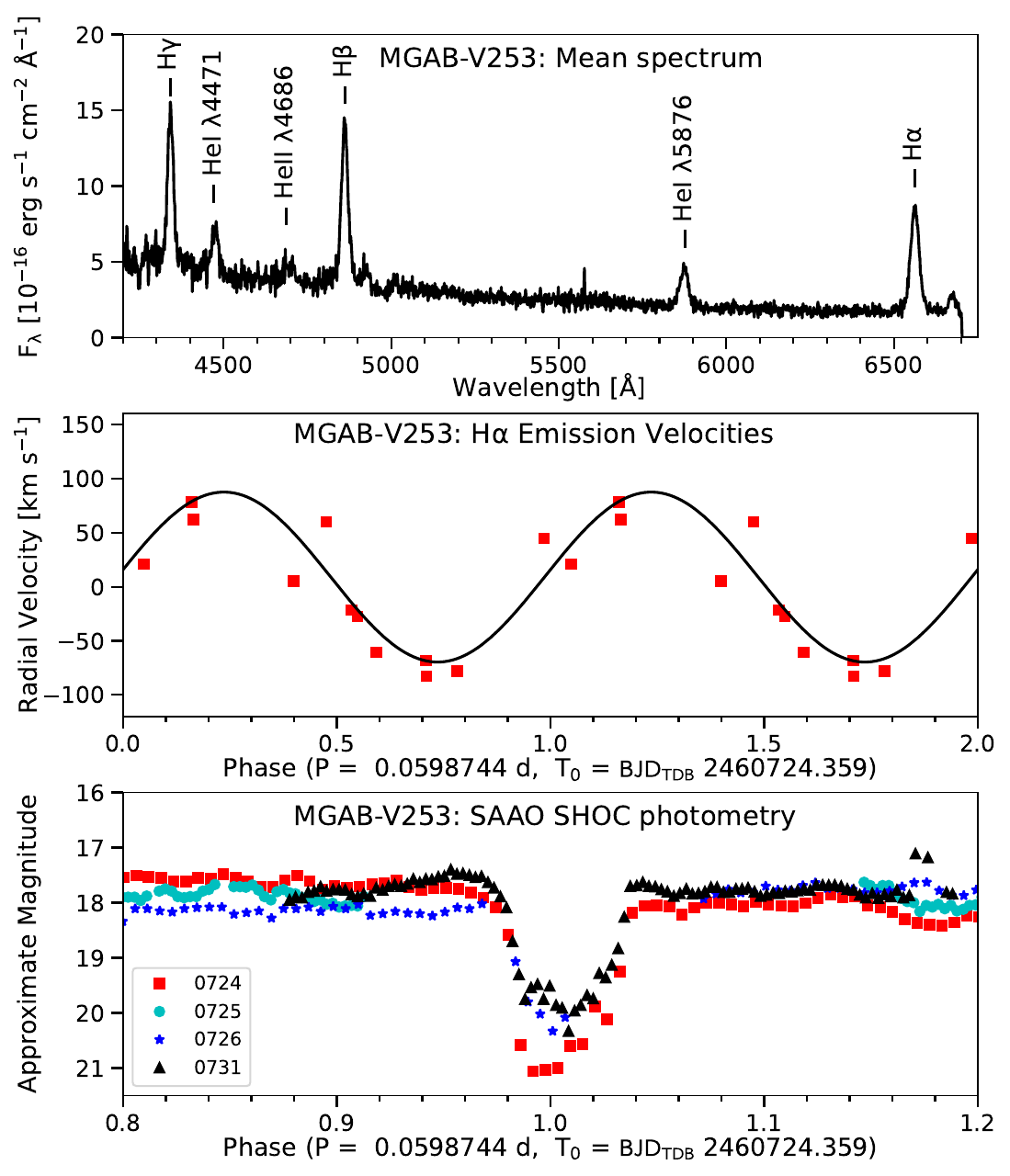}
\caption{
{\it Upper panel:} Mean fluxed spectrum of MGAB-V253 from 2025-02-25. 
{\it Middle panel:} Averaged H$\alpha$ and H$\beta$ emission-line velocities
folded on the orbital period, with the best-fit sinusoid superposed.
{\it Lower panel:} Time series photometry from the SAAO 1m and SHOC 
as a function of orbital phase, in the vicinity of the eclipse.  The
different symbols are from different nights, coded in the legend by the 
last four digits of their Julian dates. 
}
\label{fig:mgab253_spec_vcurve_shoc} 
\end{figure}

Since $P_{\rm orb}$ was known from eclipses, we did not attempt to
determine an independent radial velocity period, and obtained just
over one orbit of spectroscopy.
The top panel of Fig.~\ref{fig:mgab253_spec_vcurve_shoc} shows the average
fluxed spectrum.
The emission lines are notably
strong; H$\alpha$ and H$\beta$ have emission equivalent widths near 125 and
84 \AA\ respectively.  They are also broad ($\sim 30$ \AA\ Gaussian FWHM for
H$\alpha$), and single-peaked.  
%The eclipse requires a high inclination, 
%so the breadth is expected.  High-inclination novalikes variables generally
%show single-peaked emission, while short-period, high-inclination dwarf
%novae tend to show two-horned profiles from an optically thin disk.
% This, along with the historical light curve, corroborates the classification
% of this object as a novalike variable.  
HeII$\lambda$ 4686 is detected, but rather weakly, with an emission 
equivalent width around 10 \AA .  
The middle panel of Fig.~\ref{fig:mgab253_spec_vcurve_shoc} shows the radial 
velocities of H$\alpha$ folded on the known $P_{\rm orb}$; the 
modulation is readily apparent.  The phase, however, is unexpected,
in that the blue-to-red crossing occurs near the eclipse phase
(the ephemeris is discussed below).  This is almost diametrically
opposite the case in which the emission line radial velocities follow 
the motion of the white dwarf.  We do not have a ready explanation.
and we do not have enough phase redundancy or spectral resolution
to generate a useful phase-binned spectrum.

Our time-series photometry covered two full eclipses, and half of another.
The lower panel of Fig.~\ref{fig:mgab253_spec_vcurve_shoc} shows
the light curves near eclipse.  The apparent variation of the eclipse depth from 
one eclipse to another is likely spurious, since the object became
too faint to measure accurately.  The ingress and egress are 
steep, but not near-instantaneous as is sometimes seen in 
polars/AM Her stars.  

\begin{figure} 
\vspace*{-1cm}\hspace*{-0.0cm}\includegraphics[width=6.5 truein]{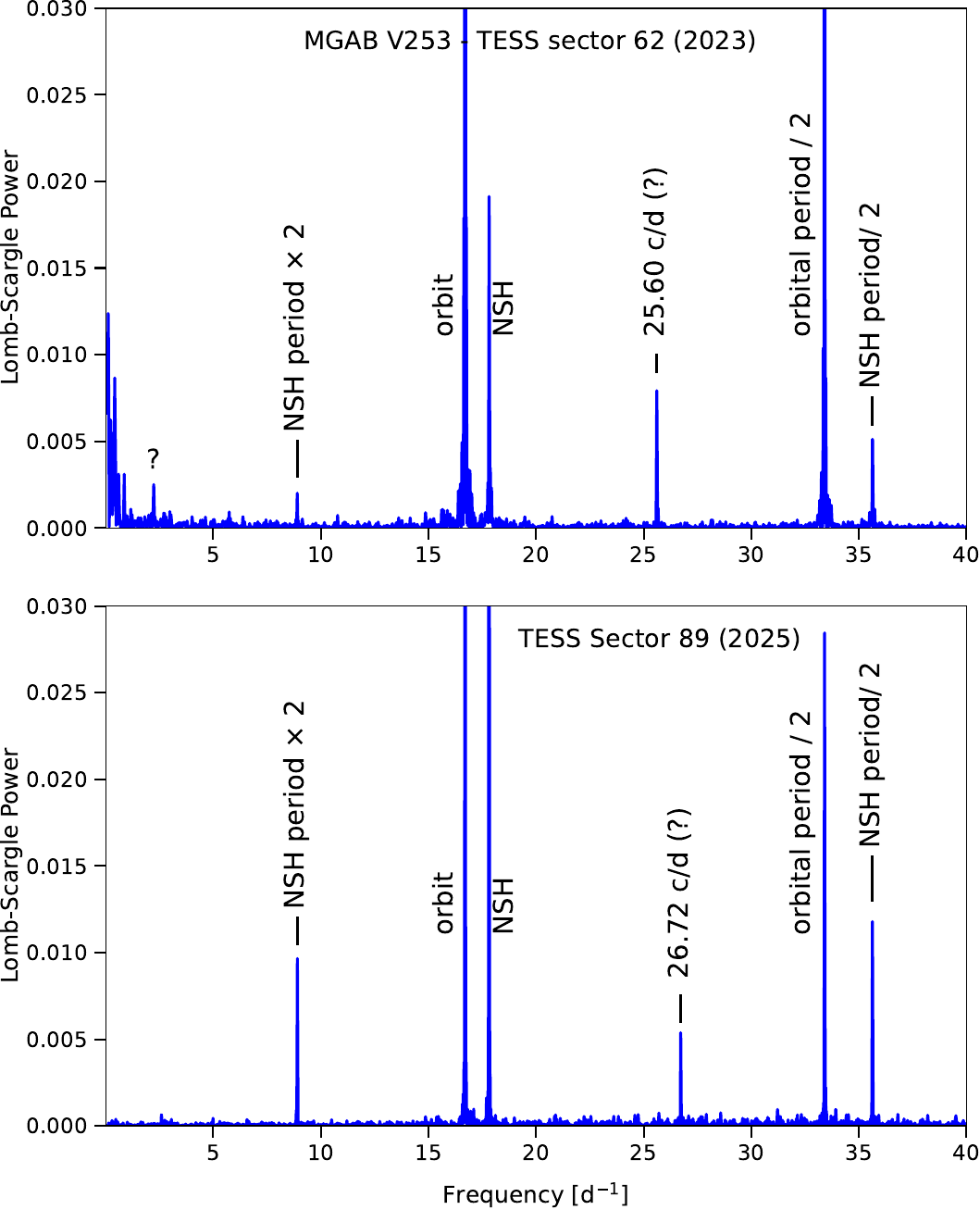}
\caption{
{\it Upper panel:} Periodogram of the Sector 62 TESS data.  The vertical
scale is set to show subtle features, leaving the peaks at $P_{\rm orb}$ and $P_{\rm nsh}$ 
off-scale, 
{\it Lower panel:} Similar to the upper panel, for Sector 89.
}
\label{fig:mgab253pgrmplot} 
\end{figure}

\begin{figure} 
\vspace*{-1cm}\hspace*{-0.0cm}\includegraphics[width=6.5 truein]{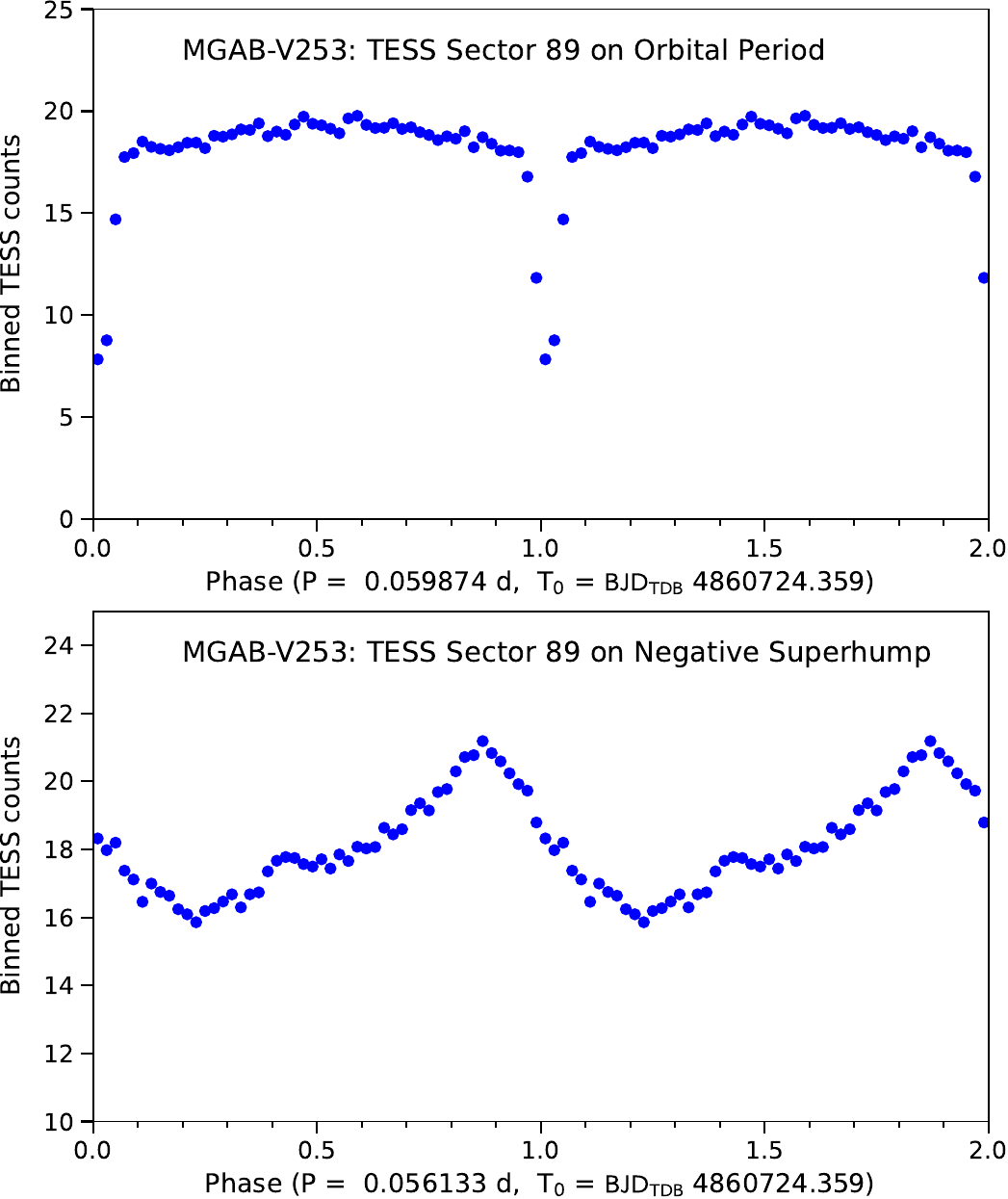}
\caption{
{\it Upper panel:} Periodogram of the Sector 89 TESS data binned on the orbital period.
{\it Lower panel:} Periodogram of the Sector 89 TESS data binned on the negative superhump period.
}
\label{fig:mgab253tessbinned} 
\end{figure}

TESS observed MGAB-V253 in Sectors 62 (2023) and 89 (2025).  
Fig.~\ref{fig:mgab253pgrmplot} shows the Lomb-Scargle periodograms
of these two sectors.  Both show prominent signals at $P_{\rm orb}$, 
but also at a slightly shorter period, near
$P = 0.05613$ d, which we identify as a negative superhump.  These are thought
to be related to the precession of a tilted accretion disk.  
Other features are harmonics of $P_{\rm orb}$ and $P_{\rm nsh}$, 
except for weak periodicities near 25.60 and 26.72 d$^{-1}$
in Sectors 62 and 89 respectively.  Fig.~\ref{fig:mgab253tessbinned}
shows the TESS measurements from Sector 89 phase-binned on 
$P_{\rm orb}$ and $P_{\rm nsh}$; the Sector 62 appear similar. 
We tried phase-binning on the curious $\sim 26$ c d$^{-1}$
periodicities as well, but found no convincing modulation, and
have no explanation for their appearance in the periodogram.

{\it Ephemeris.}  To confirm and refine the ephemeris, we started with 3887 
usable measurements from ATLAS spanning from 2015 December to 2025 February. 
These defined an unambiguous eclipse period near 0.05987440(5) d.
Next, we assembled eclipse timings from our SHOC data and the TESS sectors.
In addition to the two publicly-available TESS sectors,
Dr.~Christina Hedges of the NASA TESS Science Support Office graciously provided
data from Sector 35, extending the time base of the TESS observations.  
Our best ephemeris for
mid-eclipse is 
$$T = {\rm BJD (TDB)}\ 2460724.3593(3) + 0.05987438(3) E,$$
where $E$ is the integer cycle count.  The quoted epoch corresponds
to an eclipse observed with SHOC.  The period is nearly the same as
Murawski's initial estimate, but slightly longer. 

\begin{figure} 
\vspace*{-1cm}\hspace*{-0.0cm}\includegraphics[width=6.5truein]{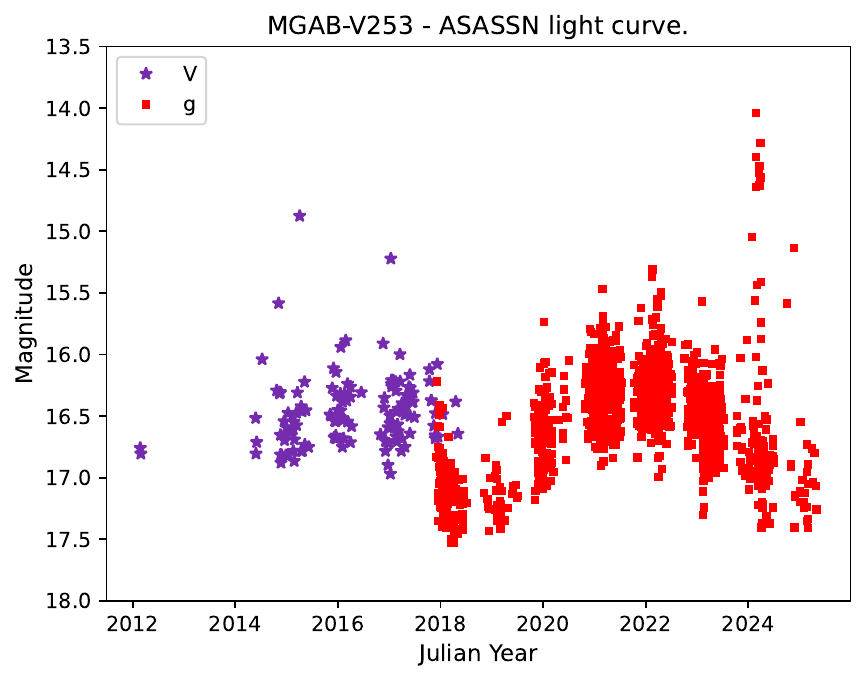}
\caption{
Light curve of MGAB-V253 from ASAS-SN. .
}
\label{fig:mgab253asassnlc} 
\end{figure}

{\it What type of CV is MGAB-V253?} Fig.~\ref{fig:mgab253asassnlc} shows
the ASASS-SN light curve of MGAB-V253.  A fold of these data on the 
eclipse ephemeris (not shown) shows that essentially none of the points 
were taken in eclipse, evidently because the eclipses dropped below
ASAS-SN's detection limit.  The long, slow variations over years are
obvious, but there is only one event (in 2024) that resembles a dwarf nova 
outburst.  

\begin{figure} 
\vspace*{-1cm}\hspace*{-0.0cm}\includegraphics[width=6.3 truein]{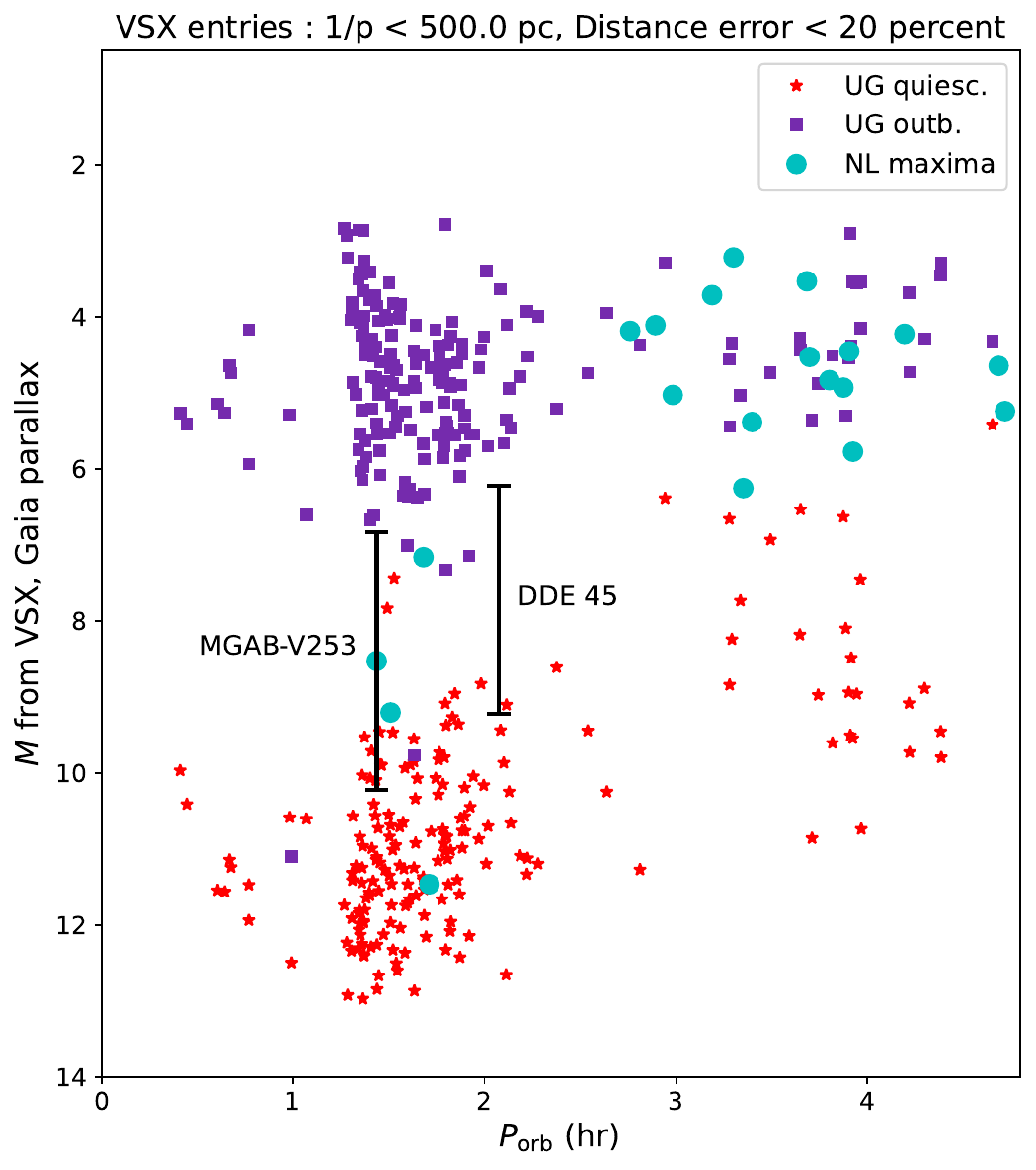}
\caption{
Absolute magnitudes of nearby CVs as a function of period and (for dwarf
novae) outburst state.  See the text details.
}
\label{fig:absmagtype-fsp25} 
\end{figure}

To put this object in context, we prepared Fig.~\ref{fig:absmagtype-fsp25},
which shows the absolute magnitudes of CVs selected from the VSX catalog plotted versus $P_{\rm orb}$.  
The sample is limited to objects with orbital periods quoted in the 
VSX, with Gaia DR3 parallax errors of less than 20 per cent,
and with $1/p < 500$ pc.  The ``novalike'' objects are those 
with VSX variability types that included the string ``NL''.
The absolute magnitudes of the novalikes are computed using their 
brightest VSX magnitudes and nominal Gaia parallaxes.  
Some novalikes, known as VY Sculptoris stars, fade irregularly by several 
magnitudes; we exclude those faint states by showing only the bright-state 
magnitudes.
We made no attempt to correct for reddening, which should mostly be modest at 
these rather short distances. MGAB-V253 is included among the novalikes; 
the vertical bar is not an uncertainty estimate, but indicates the range of 
variation found from the ASAS-SN photometry (Fig.~\ref{fig:mgab253asassnlc}).  
To select the dwarf nova sample, we included objects
with "UG" in the classification string, but {\it without} an "E" indicating
a known eclipse, so that the quiescent magnitudes are not
distorted by eclipses.  For the UGs (which include Z Cam stars, or 
UGZs), we show both the outburst and quiescent magnitudes.

There is much to dispute in this diagram -- in particular, we did
not vet any of the VSX data, and we also make no claims of 
completeness.  Nonetheless, there are some striking features.  
The period minimum near 1.25 hours is sharply defined; CVs with 
shorter periods tend to be double degenerates.  Dwarf novae
in outburst almost all have $M \sim 4.3 \pm 1$, independent
of $P_{\rm orb}$, but their quiescent magnitudes are clearly
correlated with $P_{\rm orb}$.  The bright-state absolute magnitudes of novalikes on the longward
side of the 2-3 h period `gap' \citep{knigge11} are similar to the absolute magnitudes of outbursting dwarf novae, but 
at short periods, novalikes are {\it less} 
luminous than outbursting dwarf novae at comparable periods
\footnote{The single apparent novalike among the quiescent
short-period dwarf novae traces to a questionable classification.
It is the very nearby `post-bounce' candidate GD 552
\citep{unda-sanzana08}, which appears to be physically 
similar to UGWZ systems at 
minimum light, but which has never been observed to outburst.}.

MGAB-V253 is distinctly fainter than outbursting dwarf novae, and
even at its brightest barely reaches a comparable brightness.
Thus, MGAB-V253's absolute magnitude places it among 
the novalikes -- Murawski's original classification appears accurate.
Its spectrum is also more typical of a novalike than a quiescent
dwarf nova, and it shows persistent negative superhumps. 
%, shortward of the so-called 2-3 hour
%`gap' in the CV period distribution 

% MGAB-V253 appears similar in many ways to EX Hya, a very nearby (57 pc 
% according to Gaia DR3), eclipsing intermediate polar.  Their orbital periods
% are respectively 1.44 and 1.64 hours, shortward of the so-called 2-3 hour
% `gap' in the CV period distribution \citep{knigge11}.   
% The optical spectrum of EX Hya published by \citet{williams83} closely
% resembles that of MGAB-V253.

\subsection{DDE 45} 

This variable star was discovered by Denis Denisenko, and is classified
as `CV' in his on-line list\footnote{\url{https://scan.sai.msu.ru/~denis/VarDDE.html}}.
He describes it in VSX as having an ``Unusual light curve with 3 states: 
low quiescence, mid-quiescence $\sim 1$m brighter and outbursts.''
In vsnet-chat 8129, Kato classifies it as a Z Cam star. 

\begin{figure} 
\vspace*{-1cm}\hspace*{-0.0cm}\includegraphics[width=6.3 truein]{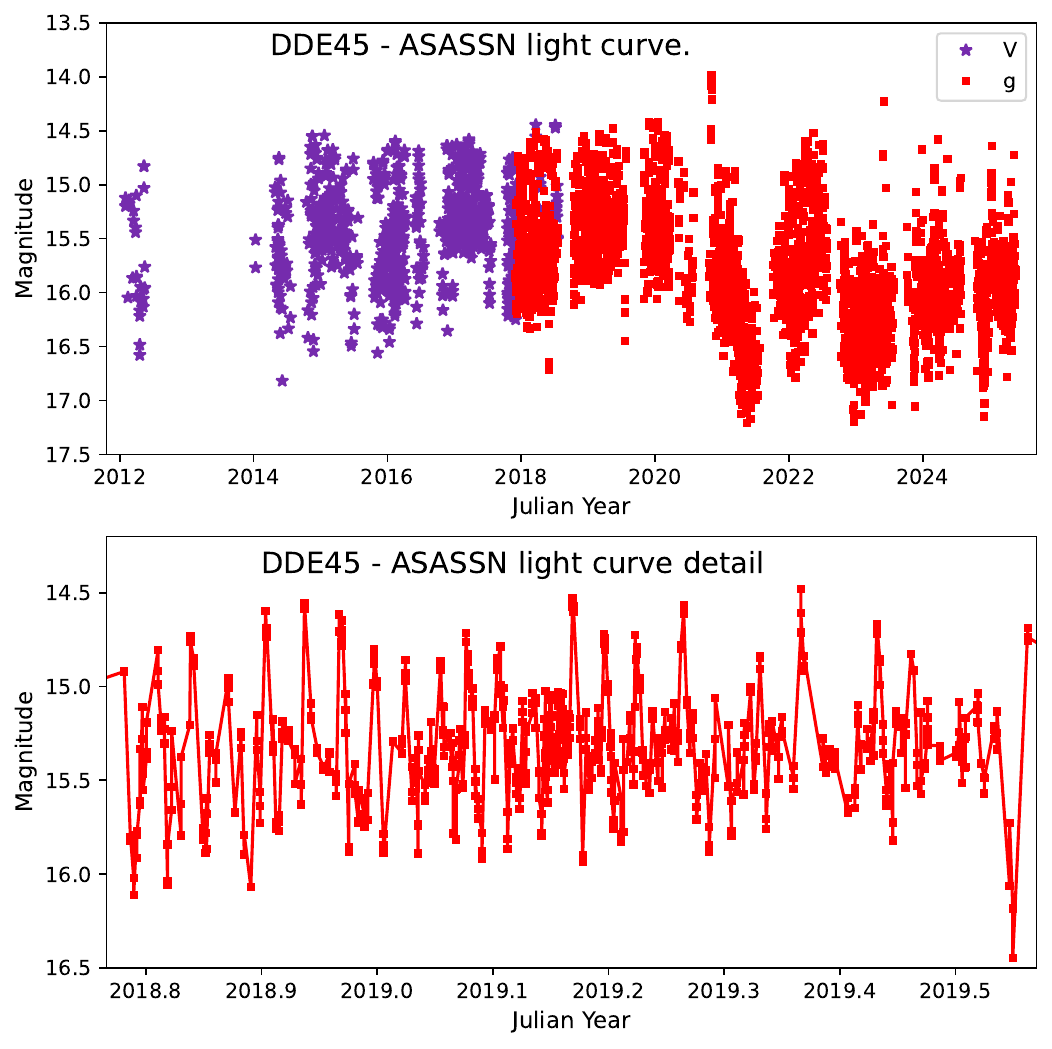}
\caption{
ASASSN light curve of DDE 45.  The lower panel shows the 2018-2109 season in 
isolation and includes connecting lines added to bring out the pattern of variability.
}
\label{fig:dde45-asassn-plot}
\end{figure}

The ASASSN light curve (Fig.~\ref{fig:dde45-asassn-plot}) shows a relatively
high state from the beginning of observations until around 2021.  The lower
panel gives a magnified view of this high state; it can be seen that most of
the variation is in outbursts and fadings on a time scale of about 10 days,
suggesting a dwarf nova with a very short outburst cycle.
characteristic of an ER UMa-type dwarf nova.  It then faded somewhat, recovered
in 2022, and since then resembles a U Gem star with a low outburst 
amplitude ($\sim 1.5$\ mag).  

%The inverse of the Gaia DR3 parallax is 359 pc; we used this together 
%with the ASASSN light curve to add 

\begin{figure} 
\vspace*{-1cm}\hspace*{-0.0cm}\includegraphics[width=6.3 truein]{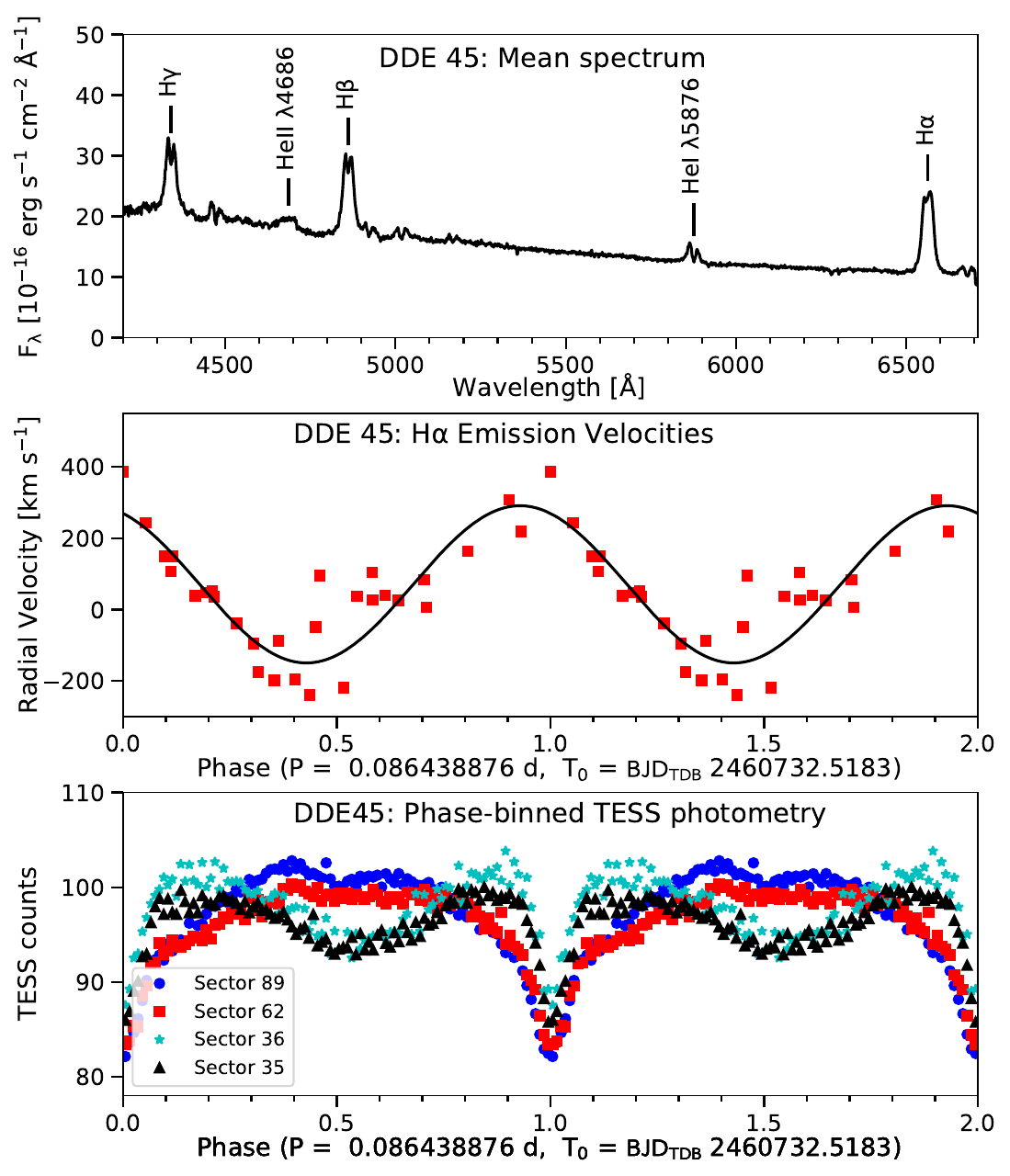}
\caption{
{\it Top panel:} Mean fluxed spectrum of DDE 45, with emission lines identified.
{\it Middle panel:} Radial velocities of the H$\alpha$ emission line 
together with a fitted sinusoid, folded on the adopted orbital period.  
The ephemeris used for the fold is the long-term eclipse ephemeris from TESS. 
{\it Lower panel.} Data from the four available TESS segments folded 
on the adopted ephemeris and averaged into 100 phase bins. 
}
\label{fig:dde45-spec-vcurve-tess}  
\end{figure}

Our mean spectrum (Fig.~\ref{fig:dde45-spec-vcurve-tess}) shows 
double-peaked Balmer and HeI emission lines. The emission lines are well-defined, 
but their equivalent widths are weaker than usual for
a quiescent dwarf nova;
H$\alpha$ has an emission equivalent width of $\sim50$ \AA .  The lines are
broad (the FWHM of H$\alpha$ is 1800 km s$^{-1}$)
and double-peaked, with peaks separated by $\sim 1000$ km s$^{-1}$.   
The radial velocities of H$\alpha$ unambiguously determine a period near 
0.08646 d (2.07 h); Fig.~\ref{fig:dde45-spec-vcurve-tess} shows
the velocities as a function of the phase defined below. 
Note that the period puts DDE 45 close to the low end of the
2-3 hour period `gap'.  

We downloaded the 120-s TESS SPOC light curves from Sectors 35 and 36
(2021), Sector 62 (2023), and 89 (2025); the last was nearly contemporaneous
with our spectroscopy.  Lomb-Scargle periodograms of latter two 
data sets showed a strong periodicity consistent with the radial 
velocity period, and folding the data from the two segments 
revealed an eclipse, along with a broad single-humped orbital 
modulation (Fig.~\ref{fig:dde45-spec-vcurve-tess}, lower panel).  
Folding the data from the earlier segments (35 and 36) on the
preliminary period clearly showed the eclipse, in addition to a
double-peaked orbital modulation, in contrast to the single
maximum seen in the later segments.
The full duration of the eclipse is about 
13 per cent of the orbital period.  Again, the modulation
was barely visible in folds of the original data, but became
beautifully clear on phase-binning.

The eclipse is not evident in the ASASSN or 
Atlas light curves, but the TESS data define a nearly
unambiguous long-term ephemeris,
%\footnote{The 8500-cycle interval between the first two sectors
%and the middle one could be off by one either away, and 16900-
%cycle interval between the middle and end off by two, but these
%alternatives the the 35-to-36 interval poorly.},
implying 
$$t = {\rm BJD_{TDB}} = 2460732.5183(5)  + 0.08643888(4) E$$
for mid-eclipse.  The largest residual is 31 seconds, 
and the epoch given is chosen to be contemporaneous with the spectroscopy.
Frequencies differing by one part in 8500 are
possible but unlikely.

\begin{figure} 
\vspace*{-1cm}\hspace*{-0.0cm}\includegraphics[width=6.3 truein]{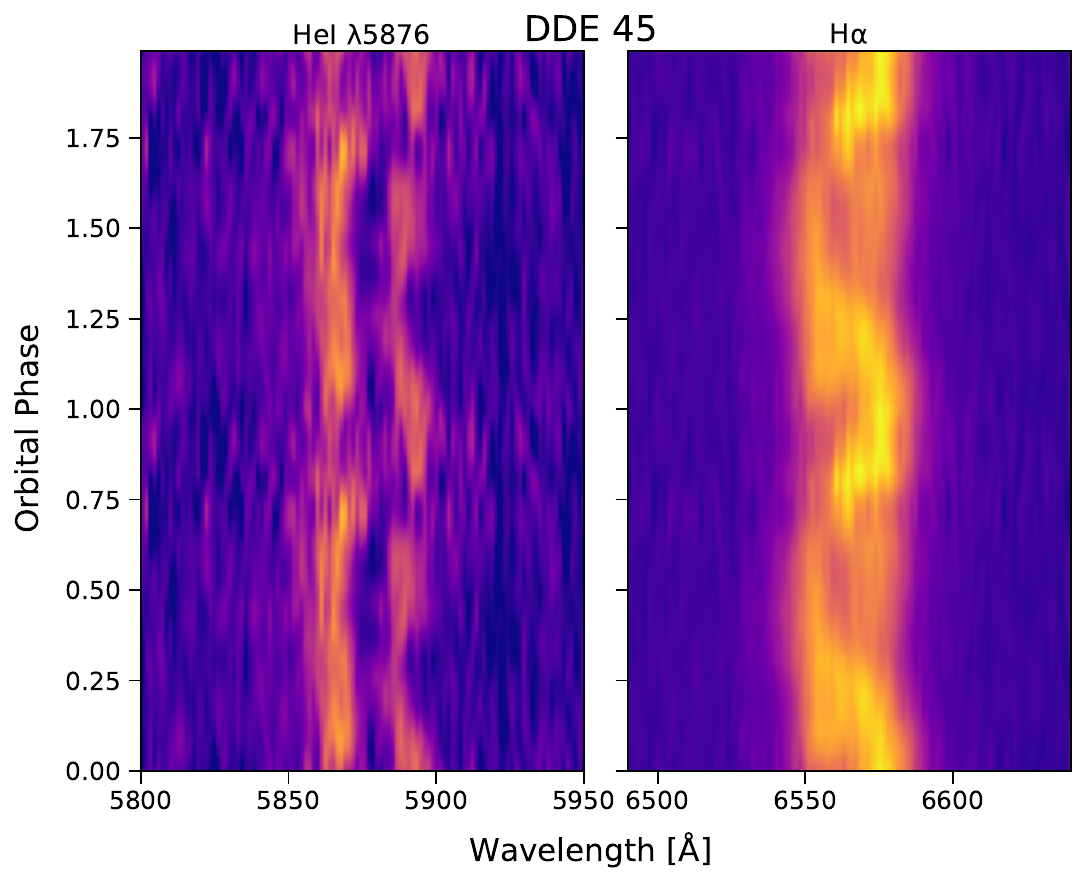}
\caption{Phase-binned spectra of DDE 45 near HeI$\lambda$5876 and
H$\alpha$.  Note the double-peaked profiles in both, and the 
indication for an S-wave component in H$\alpha$.
}
\label{fig:dde45trail}  
\end{figure}

We used this ephemeris to prepare a phase-averaged spectrum; 
Fig.~\ref{fig:dde45trail} shows the HeI $\lambda$ 5876 and 
H$\alpha$ lines, both of which show similar double-peaked 
profiles indicating of an accretion disk, along with an 
S-wave snaking back and forth between the peaks. The S-wave
appears to pass maximum positive velocity around the time
of eclipse.  If it arises at the point in which the mass-transfer
stream strikes the disk, it would be expected to have a 
slightly later phase, but it is not clear from these
data that the apparent offset is significant. 

We plot DDE 45 on Fig.~\ref{fig:absmagtype-fsp25}, 
using a distance of 359 pc and an apparent magnitude range
of 14.0 to 17.0, derived from the ASASSN light curve. 
It apparently does not fade as deeply in quiescence
as other U Gem stars at comparable periods, and is also
atypically faint at peak brightness. While it resembled
a U Gem star at the epoch of our observations, it appears
that its long-term mass-transfer rate is larger than that of a 
typical U Gem star, but somewhat less than a Z Cam
star.  The shift of variability patterns seen around 
2020-2022 in Fig.~\ref{fig:dde45-asassn-plot} suggests a
modest secular decrease in the mass-transfer
rate. 

\section{Concluding Remarks}

Table~\ref{tab:partab} summarizes the orbital parameters determined
in this paper.

%\begin{rotatetable}
\begin{deluxetable*}{llrlrrr}
\label{tab:partab}
\tablecolumns{7}
\tablewidth{0pt}
%\tabletypesize{\scriptsize}
\tablecaption{Orbit Parameters}
\tablehead{
\colhead{Object} & 
\colhead{$P_{\rm orb}$} &
\colhead{$P$ Source} &
\colhead{$T_0$\tablenotemark{a}} &
\colhead{$T_0$ Source} &
\colhead{$K$} &
\colhead{$\gamma$} \\
\colhead{} &
\colhead{(d)} &
\colhead{} & 
\colhead{BJD-TDB} &
\colhead{} &
\colhead{(km s$^{-1}$)} &
\colhead{(km s$^{-1}$)} \\
}  % end of tablehead
\startdata
6dF0752-54 & 0.210640(6) & T,AT Ec & 2460728.290(4) & RV & 86(11) & 45(7) \\
J0916-26 & 0.1404946(1) & AT,S ,Ec  & 2457424.7785(17) & S, Ec & \nodata & \nodata\\
GSC08944 & 0.158265(2) & RV,T,AS & 2460720.347(4) & RV & 93(15) & 29(10) \\
MGAB-V253 & 0.05987438(3) & T,AT,S Ec & 2460724.3593(3)\tablenotemark{b} & Ec & 78(8) & 9(6) \\
DDE 45 & 0.08643888(4)\tablenotemark{c} & T Ec & 
   2460732.5183(5)\tablenotemark{d} & Ec & 220(32) & 71(22) \\
\enddata
\tablecomments{Codes for sources of $P$ (column 3) and $T_0$ (column 5); 
T = TESS; AT = Atlas; AS = ASASSN; S = SHOC; Ec = eclipse; RV = radial velocity fit.}
\tablenotetext{a}{Radial velocity fits are of the form $v(t) = \gamma + K \sin[2 \pi (t - T_0) / P]$.} 
\tablenotetext{b}{The blue-to-red crossing of the RV curve was at 2460722.501(1),
which is at phase 0.98 in the eclipse ephemeris given here.}
\tablenotetext{c}{There is a small possibility of a cycle count error between TESS
segments. The uncertainty given here assumes the count is accurate.}
\tablenotetext{d}{The blue-to-red crossing of the RV curve occurred at 2460728.342(2), 
which is at phase 0.68 in the eclipse ephemeris given here.}
\end{deluxetable*}

The objects presented here were selected primarily for tractability, in 
particular the likelihood of obtaining concrete results in a single 
observing run using meter-class telescopes.  They illustrate the 
rich variety of CV behaviors, including a longer-period dwarf nova 
(6dF 0752-54), 
an apparently magnetic system that eclipses (J0916$-$26), an 
SW-Sex-like novalike (GSC 08944), and two somewhat unusual
short-period systems, one of them resembling a novalike variable
(MGAB V253).  

As wide-field surveys improve -- in depth, wavelength coverage, and 
time sampling -- it is becoming possible to compile volume-limited
samples of cataclysmics, finally allowing meaningful comparison
with population-synthesis models.  For example, \citet{pala20} 
compiled a nearly complete list of CVs within 150 pc, and 
more recently \citet{rodriguez25} took steps toward extending
the completeness limit to 500 pc and eventually 1 kpc.  Small,
focused studies (such as this one) that characterize systems
in detail complement these efforts.  Firmly-established
orbital periods, classifications, and so on, enable richer 
analyses of these remarkable samples.
A happy (if not entirely intentional) consequence of our
selecting brighter targets out of observational necessity is
that, as Table \ref{tab:starinfo} shows, two of our small sample lie 
comfortably within 500 pc, and all within 1 kpc.

\begin{acknowledgments}

This paper uses observations made at the South African
Astronomical Observatory (SAAO),
The SAAO observations were taken as part of the 2025 Dartmouth
Foreign Study Program in Astronomy.  
%We thank Professor Ryan
%Hickox and graduate student Emmanuel Durodola for supervising
%the time-series photometry measurements.  
In addition to the
authors of this paper, Dartmouth undergraduates
Aryan Bawa, Michael Farnell, % Shreya Gandhi, 
Piper Gilbert,
% Gavin Goss, 
Kushal Jayakumar, Grace Kallman, Alexandra Lipschutz, Timothy McGrath, 
Ricardo Mendez, 
% Annabelle Niblett, 
Kate Schwendemann,
Beatrice Sears, 
%Arnav Singh, 
Aimilia Tsopela, 
% Divik Verma,
and Kendall Yoon
%, and Lauren Zanarini 
assisted.

We thank Christina Hedges and Jason Eastman for useful conversations regarding
time systems, and especially Dr.~Hedges for assistance with the TESS data. 

This work has made use of data from the Asteroid Terrestrial-impact Last Alert
System (ATLAS) project. The Asteroid Terrestrial-impact Last Alert System
(ATLAS) project is primarily funded to search for near earth asteroids through
NASA grants NN12AR55G, 80NSSC18K0284, and 80NSSC18K1575; byproducts of the NEO
search include images and catalogs from the survey area. This work was
partially funded by Kepler/K2 grant J1944/80NSSC19K0112 and HST GO-15889, and
STFC grants ST/T000198/1 and ST/S006109/1. The ATLAS science products have been
made possible through the contributions of the University of Hawaii Institute
for Astronomy, the Queen’s University Belfast, the Space Telescope Science
Institute, the South African Astronomical Observatory, and The Millennium
Institute of Astrophysics (MAS), Chile.

This paper includes data collected by the TESS mission.
Funding for the TESS mission is provided by the NASA's Science
Mission Directorate.  

This work has made use of data from the European Space Agency (ESA) mission
{\it Gaia} (\url{https://www.cosmos.esa.int/gaia}), processed by the {\it Gaia}
Data Processing and Analysis Consortium (DPAC,
\url{https://www.cosmos.esa.int/web/gaia/dpac/consortium}). Funding for the DPAC
has been provided by national institutions, in particular the institutions
participating in the {\it Gaia} Multilateral Agreement.

Finally, we would like to thank the referee for a thoughtful
and careful report. 

\end{acknowledgments}

\bibliographystyle{aasjournalv7}
\bibliography{ref}

\end{document}